\documentclass[pra,twocolumn,superscriptaddress,nofootinbib,noshowpacs,preprintnumbers,longbibliography,floatfix]{revtex4-2}
\usepackage[utf8]{inputenc}
\usepackage[english]{babel}
\usepackage{graphicx}
\usepackage{float}
\usepackage{amssymb}
\usepackage{amsmath}
\usepackage{dsfont}
\usepackage{array}
\usepackage[normalem]{ulem}
\usepackage{bm,fixmath}
\usepackage[usenames, dvipsnames]{xcolor}
\usepackage{physics}
\usepackage{tikz}
\usepackage{qcircuit}
\usepackage{booktabs}
\usepackage{placeins}
\usepackage[pdftex,
            pdftitle={QSE for wave packet generation in a Z2 LGT},
            pdfauthor={Yahui Chai, Yibin Guo, Stefan Kuehn},
            bookmarks,
            colorlinks,
            linkcolor=myblue,
            citecolor=mymagenta,
            menucolor=black,
            urlcolor=myblue,
            plainpages=false,
            pdfpagelabels,
            hypertexnames=false]{hyperref}
\usepackage{orcidlink}
\usepackage{comment}
\graphicspath{{figure/}}

\definecolor{mymagenta}{RGB}{200, 0, 100}
\definecolor{myblue}{RGB}{45, 48, 146}

\begin{document}
\title{Scalable Quantum Algorithm for Meson Scattering in a Lattice Gauge Theory}

\author{Yahui Chai\orcidlink{0000-0002-3404-6096}}
\affiliation{Deutsches Elektronen-Synchrotron DESY, Platanenallee 6, 15738 Zeuthen, Germany}
\email{yahui.chai@desy.de}

\author{Yibin Guo\orcidlink{0000-0003-0435-1476}}
\affiliation{Deutsches Elektronen-Synchrotron DESY, Platanenallee 6, 15738 Zeuthen, Germany}

\author{Stefan Kühn\orcidlink{0000-0001-7693-350X}}
\affiliation{Deutsches Elektronen-Synchrotron DESY, Platanenallee 6, 15738 Zeuthen, Germany}

\date{\today}

\begin{abstract}
    Scattering processes are fundamental for understanding the structure of matter, yet simulating their real-time dynamics remains challenging for classical computers. Quantum computing and quantum-inspired methods offer a promising avenue for efficiently simulating such phenomena. In this work, we investigate meson scattering in a (1+1)-dimensional $\mathds{Z}_2$ lattice gauge theory with staggered fermions. We develop a quantum subspace expansion technique to construct high-fidelity meson creation operators across a broad range of masses and momenta.
    Using Tensor Networks simulations, we study both elastic and inelastic scattering and provide a detailed analysis of energy transfer, entanglement entropy, and new particle production during the dynamics. In addition, we design an efficient quantum circuit for meson wave packet preparation using Givens rotations, significantly reducing the circuit depth compared to existing methods. Our work provides a non-variational and scalable framework for simulating meson scattering on near-term quantum devices, and provides a concrete strategy for quantum simulation to analyze non-perturbative dynamical processes in confining gauge theories. 
\end{abstract}

\maketitle

\section{Introduction}
Scattering processes are central to our understanding of fundamental interactions and the structure of matter. Particle collision experiments, such as those at the Large Hadron Collider (LHC)~\cite{Bruning:2012zz}, probe theoretical predictions and search for physics beyond the Standard Model. However, simulating the real-time dynamics of scattering from first principles remains a major challenge. This is because the conventional numerical approach to gauge field theories, which relies on discretizing the Lagrangian on a Euclidean spacetime lattice, is not directly applicable to dynamical processes. While this formulation enables efficient Monte Carlo methods and has proven successful for static observables~\cite{Durr2008, Alexandrou_2014}, it suffers from the sign problem when extended to real-time evolution in Minkowski spacetime~\cite{pan2022sign}, as the probability weights become complex and prevent efficient sampling. As a result, many key aspect such as the intermediate states or the evolution of quantum correlations during the scattering process are poorly understood, as they are also difficult to probe experimentally. 

Consequently, there is an ongoing search for alternative methods that can overcome these limitations. In particular, quantum-inspired methods, such as
Tensor Network States (TN), provide a promising alternative for addressing regimes inaccessible with conventional Monte Carlo methods. While TN has already been successfully demonstrated to overcome the sign problem~\cite{Banuls2018a,Banuls2019,silvi_finite-density_2017,magnifico_lattice_2021,angelides_computing_2023}, they are only efficient in situations where the entanglement is moderate. In out-of-equilibrium processes like particle collisions, entanglement can grow rapidly, restricting TN simulations to short time scales~\cite{Calabrese_2005,Schuch_2008,Schachenmayer_2013}. In addition, although meson scattering has been simulated with TN in (1+1)-dimensional gauge models~\cite{Rigobello_2021, Papaefstathiou_2024}, those studies rely on analytical knowledge specific to certain coupling regimes to approximate meson creation operators. A general method to construct such operators across all coupling strengths with high fidelities is still lacking, and the resource scaling remains unclear.

Quantum computing offers a promising path forward for simulating dynamical problems, as its efficiency is not limited by the maximum amount of entanglement. Several proof-of-principle quantum simulations have been realized in simple quantum field theories and lattice gauge models~\cite{Martinez2016,Klco2018,Atas2022,Surace_2021,Guoxian_2024,Farrell_2024_dynamics,chai2024, Alexandrou2025, Gustafson_2021, Mildenberger_2025, Zhou_2022}. However, implementing scattering processes on a digital quantum computer poses specific challenges, in particular preparing particle wave packets, which generally involve non-unitary particle creation operators, and cannot be directly realize by unitary quantum gates. Reference~\cite{chai2024} has addressed this issue for fermion scattering through circuit decompositions based on Givens rotation. For composite particles such as mesons, wave packet preparation is more challenging. Recent works proposed approaches for this by Trotter-based and SVD-inspired circuits~\cite{Davoudi_2023,Davoudi_2024}, or variational methods such as SC-ADAPT-VQE~\cite{Farrell_2024, Farrell_2024_dynamics, Farrell2025, zemlevskiy2024}. However, these approaches require either circuit approximations or the variational optimization of parameters, which is costly in terms of measurements required. A scalable and accurate method for preparing meson wave packets by quantum circuits is still missing.

In this work, we address these two main challenges in simulating meson scattering on quantum computers: constructing suitable meson creation operators and designing efficient quantum circuits for wave packet preparation. Using the (1+1)-dimensional $\mathds{Z}_2$ lattice gauge theory~\cite{D.Horn_1979, Kogut_1975, Susskind_1977} as a testbed, we develop a general and scalable framework. First, we introduce a quantum subspace expansion (QSE) method, which allows systematic construction of high-fidelity meson creation operators across a wide parameter range. This approach avoids variational optimization, is resource-efficient, and respects the symmetries of the theory, producing operators with well-defined quantum numbers. To study the full scattering process, we use MPS to simulate Trotterized quantum circuits, analyzing energy transfer, entanglement growth, flux string formation and breaking, and new particle production in inelastic scattering. These results indicate that inelastic scattering is more costly for TNS and underscore the importance of quantum approaches for this process. To support further hardware implementations, we also develop an efficient quantum circuit for the preparation of meson wave packets using Givens rotations~\cite{Jiang_2018,Kivlichan_2018}, which is accurate and shallower compared to previous methods. Our approach reduces the complexity of the CNOT gate from $\mathcal{O}(L^3)$~\cite{Davoudi_2024} to $\mathcal{O}(L^2)$, and the circuit depth from $\mathcal{O}(L^3)$~\cite{Davoudi_2024} to $\mathcal{O}(L)$, where $L$ is the number of lattice sites. This method significantly improves the feasibility of quantum simulation for meson scattering.

This paper is organized as follows. In Sec.~\ref{sec: Model}, we introduce the lattice $\mathds{Z}_2$ gauge theory and its symmetries. In Sec.~\ref{sec: meson_op}, we introduce the QSE construction of meson creation operators. Section~\ref{sec: scattering} shows our numerical results, including elastic and inelastic scattering. In Sec.~\ref{sec: circ_wp}, we propose an efficient quantum circuit for the preparation of meson wave packets and analyze its resource scaling. Finally, we summarize our results and discuss future research directions in Sec.~\ref{sec: summary}.

\section{The model}\label{sec: Model}
\begin{figure}[htp!]
    \centering
    \includegraphics[width=0.8\linewidth]{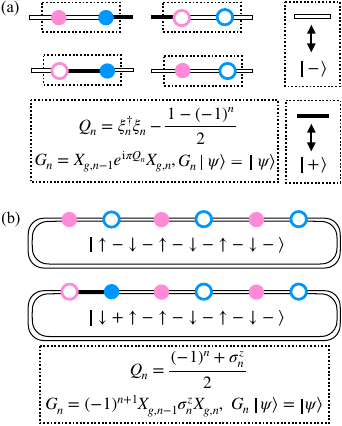}
    \caption{Illustration of one-dimensional $\mathds{Z}_2$ lattice gauge theory. (a) Example configurations satisfying the Gauss law. The pink (blue) circles correspond to odd (even) sites, and solid (empty) circles represent filled (empty) fermions. The fermions are connected by $\mathds{Z}_2$ gauge field, represented by horizontal lines. The empty (filled) line corresponds to the electric field in the eigenstate $\ket{-}$ ($\ket{+}$), with eigenvalues $-1$ ($+1$) respectively. Other valid configurations can be obtained by simultaneously flipping all gauge field qubits.
    (b) The upper part shows the ground state in the strong coupling limit $\varepsilon \rightarrow \infty$, where all electric fields take the lowest eigenvalue state, all odd (even) fermion qubits are spin up (down). A gauge link is connected between the last and first site due to the periodic boundary condition used here. The lower plot shows a meson excitation created by the operator $\xi_1 Z_{g,1}\xi_2^{\dagger}$, which annihilates a fermion at site 1 (thereby effectively creating an antifermion), and creates a fermion at site 2, resulting in an antifermion-fermion pair connected by an electric flux string to satisfy the Gauss law. 
    } 
    \label{fig:Z2LGT}
\end{figure}

In this work, we focus on a $\mathds{Z}_2$ lattice gauge theory~\cite{D.Horn_1979} coupled to staggered fermions~\cite{Kogut_1975, Susskind_1977}. It is arguably one of the simplest lattice gauge theories with fermionic matter, as the gauge degrees of freedom on each link correspond to spins with just two states. The Hamiltonian consist of three parts, $H = H_{\text{kin}} + H_{\text{mass}} + H_{\text{el}}$, where the first part corresponds to the kinetic term, the second to the mass term and the third to the electric energy term. On a lattice with $L$ matter sites, these read
 \begin{equation}\label{eq: Hamiltonian}
    \begin{aligned}
            H_{\text{kin}} &= \frac{1}{2a} \sum_{n=1}^L \left( \xi_n^{\dagger} Z_{g, n} \xi_{n+1} + \text{h.c.} \right), \\
            H_{\text{mass}} &= m \sum_{n=1}^L (-1)^n \xi_n^{\dagger} \xi_n, \\
            H_{\text{el}}&= \varepsilon \sum_{n=1}^L X_{g, n}.
    \end{aligned}
\end{equation}
In the above expression, the operators $\xi_n^{\dagger}$ ($\xi_n$) represent fermionic creation (annihilation) operators on matter site $n$, and $X_{g, n}$, $Z_{g, n}$ are the usual Pauli matrices acting on the gauge links connecting sites $n$ and $n+1$. Periodic boundary conditions are used to have a good definition of momentum, where site $L+1$ is identified with $1$ in the kinetic term. The parameters $a$, $m$ and $\varepsilon$ correspond to the lattice spacing, the fermion mass and the coupling, respectively. The gauge degrees of freedom on each link can be described by the eigenstates of the  Pauli-$X$ operator $X_{g, n}$, $\ket{+}, \ket{-}$ corresponding to electric field values $+1$ and $-1$. The physical states of the theory have to fulfill Gauss law, $\forall n\,\, G_n \ket{\psi} = \ket{\psi}$, where
\begin{equation}
    G_n = X_{g, n-1} \exp(i\pi Q_n )   X_{g, n}
    \label{eq:Z2_gauss_law}
\end{equation}
and $Q_n=\xi_n^{\dagger} \xi_n - (1-(-1)^n)/2$ is the staggered fermionic charge. Figure~\ref{fig:Z2LGT}(a) provides an illustration of the lattice system. Note that the staggered formulation essentially corresponds to separating the upper (lower) components of the Dirac spinor to even (odd) sites, hence we focus on even values of $L$ for the rest of the paper. In addition, throughout this work we target the sector with vanishing total charge, $\sum_n Q_n = 0$, which corresponds to half-filling, i.e. $\sum_n \xi_n^{\dagger} \xi_n = L / 2$. Besides, we use the dimensionless rescaling the Hamiltonian $aH$ and set $a=1$ without loss of generality. As a result, both the fermion mass $m$ and the electric field coupling $\varepsilon$ are expressed in units of the lattice spacing and are treated as dimensionless parameters. We vary $m$ and $\varepsilon$ to explore different physical regimes of the theory.

The Hamiltonian in Eq.~\eqref{eq: Hamiltonian} is invariant under charge conjugation. Due to the staggered formulation corresponding to separating the components of the Dirac spinor to different sites, the lattice version of the charge conjugation operator $C$ cannot be defined on a single staggered site, and involves translation by one lattice site~\cite{carena2024, stone2021}\footnote{The operator $C$ used here corresponds to $C_{-}$ in Ref.~\cite{carena2024}. The fermion field $\xi_n$ differs from the $\chi_n$ used in that work by a phase factor, leading to a slightly different form of the transformation in Eq.~\eqref{eq: SR} compared to Eq.~(A11) in Ref.~\cite{carena2024}, though the underlying physics is the same.}
\begin{equation}\label{eq: SR}
    \begin{aligned}
        C  \xi_n  C^{-1} &= (-1)^n \xi_{n+1}^{\dagger},\\
        C  Z_{g, n}   C^{-1} &= Z_{g, n+1}.
    \end{aligned}
\end{equation}
Note that $C^2$ corresponds to a translation by two staggered lattice sites, which leaves the Hamiltonian invariant, too, and allows for the definition of a lattice momentum. Hence, there exists an eigenbasis for the Hamiltonian in which its eigenstates have a well defined momentum and charge conjugation quantum number: $C \ket{k, c} = c e^{-ika} \ket{k, c} $. Here $c=\pm 1$ is the charge conjugation and $k$ is the lattice momentum with possible values $k \in \Lambda = (2\pi / aL) \{-L/2, -L/2+1, \cdots , L/2-1\}$. In the continuum limit, $ka \rightarrow 0$, $C$ will reduce to the usual charge conjugation operator. 

In the well-studied Schwinger model, the two stable particles are the vector and the scalar meson, which are characterized by their charge conjugation number~\cite{coleman_more_1976, banuls_mass_2013}. We observe the same behavior for the lattice $\mathds{Z}_2$ gauge theory in our numerical results: the ground state lies in the charge conjugation sector $c=+1$, and the first zero-momentum excitation is directly above the ground states with $c=-1$, hence, we refer to it as the vector meson. Similar to Ref.~\cite{banuls_mass_2013}, we observe the second zero-momentum excitation having $c=+1$, which we refer to as the scalar meson in analogy to the Schwinger model.

In order to simulate the model on a quantum computer, we map the fermionic degrees of freedom to spins using a Jordan-Wigner transformation (JWT)
\begin{equation}\label{eq: JWT}
    \begin{aligned}
        \xi_n^{\dagger} = \prod_{l<n} (-i\sigma_l^z) \sigma_n^+, \quad\quad \xi_n = \prod_{l<n} (i\sigma_l^z) \sigma_n^-,
    \end{aligned}
\end{equation}
where  $\sigma_n^\pm = (\sigma^x_n \pm i\sigma^y_n)$, and we use $\sigma^a_n$, $a=x,y,z$, for Pauli matrices acting on the matter sites to explicitly distinguish them from the gauge degrees of freedom. The resulting spin Hamiltonian reads
\begin{equation}\label{eq: Hamiltonian_spin}
    \begin{aligned}
        H &= -\frac{i}{2a} \sum_{n=1}^{L-1} \left( \sigma_n^{+} Z_{g, n} \sigma_{n+1}^- - \text{h.c.}  \right) \\
        &+ \frac{1}{2a} \left( \prod_{l<L}(-i\sigma_l^z) \sigma_L^+ Z_{g, L} \sigma_1^- + \text{h.c.} \right) \\
        &+ \frac{m}{2}\sum_{n=1}^L (-1)^n \sigma_n^z \\
        &+ \varepsilon \sum_{n=1}^L X_{g, n}.
    \end{aligned}
\end{equation}
where the second line arises due to the periodic boundary conditions and we have omitted the constant $\sum_{n=1}^L (-1)^n m/2$ in mass term, as it evaluates to zero for even values of $L$ used in this work. The Gauss law in spin language corresponds to 
\begin{align}
    G_n = (-1)^{n+1}X_{g, n-1} \sigma^z_n   X_{g, n}.
    \label{eq:Z2_gauss_law_spin_formulation}
\end{align}

Consequently, a total of $N = 2L$ qubits are required to represent both the fermionic matter fields and the gauge degrees of freedom in a system comprising $L$ sites.

\section{Meson operator construction}\label{sec: meson_op}
In the sector of vanishing total charge, because of confinement in the $\mathds{Z}_2$ gauge theory~\cite{Alexandrou2025}, we expect the excited states in this sector to be mesons, i.e., bound states of a fermion and an antifermion. This picture becomes more clear in the strong coupling limit, $\varepsilon \rightarrow \infty$, for which the contribution of the kinetic and the mass term in the Hamiltonian can be omitted. The ground state in this limit is given by
\begin{equation}
    \ket{\Omega_{\infty}} = \ket{1} \ket{-} \ket{0} \ket{-} \ket{1} \cdots \ket{-} \ket{0} \ket{-},
\end{equation}
where all gauge links arrange in the configuration minimizing the electric energy, and the odd (even) sites are occupied (empty) due to the fact that the staggered charge has to vanish to fulfil the Gauss law in Eq.~\eqref{eq:Z2_gauss_law} for this electric field configuration. Note that this state also minimizes the mass energy in the absence of the kinetic term. The excited states in this limit correspond to fermion-antifermion pairs between nearest neighbors, which must be connected by a flux tube to satisfy the Gauss law (c.f.\ Fig.~\ref{fig:Z2LGT}(b)). These excitations can be represented as
\begin{equation}
    \ket{k=0, c = -1}_{b} = \sum_n \left( \xi_n^{\dagger} Z_{g, n} \xi_{n+1} -\text{h.c.} \right) \ket{\Omega_{\infty}},
\end{equation}
where the subscript $b$ denotes a meson state, $k=0$ represents the meson's momentum and $c = -1$ specifies the charge conjugation of this meson. 

In contrast, in the weak coupling region, a meson state can be approximated as the product of zero-momentum fermions and antifermions~\cite{Rigobello_2021}. In this situation the meson has a looser structure and the fermion-antifermion pairs are connected by longer flux tubes compared to the strong coupling limit. 

For the intermediate coupling region, it is in general challenging to determine the meson operator. Reference~\cite{Davoudi_2024} proposed using a variational quantum eigensolver (VQE) to optimize the meson structure parameters to identify the lowest-energy state for a given momentum, which corresponds to the lightest meson. This method showed a good performance in the case of either large mass or a strong coupling for the system up to $L=10$. However, in practice, a VQE can be difficult to implement efficiently, especially as the system size grows~\cite{Anschuetz_2022}. In this work, we instead propose to use QSE~\cite{McClean_2017, McClean_2020, Takeshita_2020, Yoshioka_2022}, thus avoiding VQE completely. We aim at obtaining meson states with high fidelity over a wide parameter region and a wide range of momenta. Specifically, we expect the operator creating a meson with momentum $k$ and charge conjugation $c$ have the form
\begin{equation}\label{eq: bdag}
    \begin{aligned}
        b_{k, c}^{\dagger} &= \sum_{n,l=1}^{L} a_{n,l}^{(k,c)}~ \xi_n^{\dagger} W_{n, l} \xi_l \equiv \sum_{I} a_I^{(k,c)} ~ M_I
    \end{aligned}
\end{equation}
 where $a_{n,l}^{(k,c)}$ are complex coefficients. To simplify the expression in QSE, we define a multi-index $I = (n, l)$ and the shorthand operator $M_I = \xi_n^{\dagger} W_{n, l} \xi_l$. The operator $W_{n, l}$ is the Wilson line connecting the fermionic operators at sites $n$ and $l$. Since we consider periodic boundary condition, there are two possible paths to connect these two sites, where it is energetically favorable to choose the shorter one to have lower electric field energy. For $n < l$, the Wilson line is defined as
\begin{equation}\label{eq: W_nl}
    W_{n, l} = \begin{cases}
        \displaystyle \prod_{r=n}^{l-1} Z_{g, r},  & l-n \leq \frac{L}{2}, \\
        \displaystyle  \left( \prod_{r=1}^{n-1} Z_{g, r} \right) \left( \prod_{r=l}^{L} Z_{g, r} \right), & l-n \geq \frac{L}{2},
    \end{cases}
\end{equation}  
and similarly for the case $n > l$. Note that the Wilson line only involves Pauli-$Z$ operators acting on the links and, thus, is self-adjoint.

With appropriate choice of coefficients $a_{I}^{(k,c)}$, the operator $b_{k, c}^{\dagger}$ will excite a proper meson state from the ground state $\ket{\Omega}$, 
\begin{equation}\label{eq: State_kc}
        \ket{k, c}_b = b_{k, c}^{\dagger} \ket{\Omega} = \sum_{I} a_I^{(k,c)}~ M_I \ket{\Omega}
\end{equation}
which is an eigenstate of both the Hamiltonian and the charge conjugation operator
\begin{equation}\label{eq: eigen_HC}
    \begin{aligned}
        H \ket{k, c}_b &=E \ket{k, c}_b,\\
        \ C \ket{k, c}_b &= c e^{ik} \ket{k, c}_b.
    \end{aligned}
\end{equation}
To find the optimal coefficients $\vec{a}^{(k,c)} = \{a_{1,1}^{(k,c)}, a_{1,2}^{(k,c)}, \cdots \}$,  we solve the generalized eigenvalue problem
\begin{equation}\label{eq: QSE1}
    \left( \mathcal{H }+ \mathcal{C} \right) \vec{a}^{(k,c)}= \lambda ~ \mathcal{S} \vec{a}^{(k,c)},
\end{equation}
on the subspace
\begin{equation}\label{eq: QSE_basis}
    \{ M_I \ket{\Omega}| I \in \{1, 2, \cdots L \}^2  \}.
\end{equation}
In Eq.~\eqref{eq: QSE1} the generalized eigenvalues are given by $\lambda = E + ce^{-ika}$ and the matrices $\mathcal{H}$, $\mathcal{C}$ and $\mathcal{S}$ have the entries
\begin{equation}\label{eq: QSE_HCSmat}
        \begin{aligned}
            \mathcal{H}_{I , J} &= \bra{\Omega} M_I^{\dagger} H M_J \ket{\Omega}, \\
            \mathcal{C}_{I, J} &= \bra{\Omega} M_I^{\dagger} C M_J\ket{\Omega}, \\
            \mathcal{S}_{I, J} &= \bra{\Omega} M_I^{\dagger} M_J \ket{\Omega}.
        \end{aligned}
\end{equation}
In addition, the meson annihilation operator $b_{k, c}$ should satisfy $b_{k, c} \ket{\Omega} = 0$. This condition can be ensured by adding an additional term to the QSE equations, otherwise the solution of Eq.~\eqref{eq: QSE1} may yield coefficients contaminated by unwanted annihilation operators. A simple example are operators like $\tilde{b}_{k,c}^{\dagger} = b_{k, c}^{\dagger} + b_{k^{\prime}, c^{\prime}}$ where $b_{k^{\prime}, c^{\prime}}$ is an arbitrary meson annihilation operator. While $\tilde{b}_{k,c}^{\dagger}$ will still excite eigenstates of $H$ and $C$, it will generally not fulfill the condition $\tilde{b}_{k,c} \ket{\Omega} =  b_{k^{\prime}, c^{\prime}}^{\dagger} \ket{\Omega}\neq 0$. Hence, an extra term has to be added to enforce that the norm $\mathcal{N_Z} \equiv \bra{\Omega}b^{\dagger}_{k,c} b_{k,c} \ket{\Omega} $ is zero, as explained in App.~\ref{app: QSE}
\begin{equation}\label{eq: QSE_Zmat}
    \begin{aligned}
        \mathcal{Z}_{I, J} &= \bra{\Omega} M_J M_I^{\dagger} \ket{\Omega}.
    \end{aligned}
\end{equation}
and resulting the final QSE equation to satisfy all requirements
\begin{equation}\label{eq: QSE}
    \left( \mathcal{H }+ \mathcal{C} + \mathcal{Z} \right) \vec{a}^{(k,c)}= \lambda ~ \mathcal{S} \vec{a}^{(k,c)}
\end{equation}
where now $\lambda = E + ce^{-ika} + \mathcal{N_Z}$. After normalizing the coefficients to satisfy $\mathcal{N_S} \equiv \vec{a}^{(k, c) \dagger} \mathcal{S} \vec{a}^{(k,c)} = 1$, the physical quantities can be extracted separately by:
\begin{equation}
    \begin{aligned}
    E &= \vec{a}^{(k, c) \dagger} \mathcal{H} \vec{a}^{(k,c)}, \\
    ce^{-ik} &= \vec{a}^{(k, c) \dagger} \mathcal{C} \vec{a}^{(k,c)}, \\ 
    \mathcal{N_Z} &= \vec{a}^{(k, c) \dagger} \mathcal{Z} \vec{a}^{(k,c)}.
    \end{aligned}
\end{equation}
We select the vectors $\vec{a}^{(k,c)}$ with lowest values of $E + \mathcal{N_{\mathcal{Z}}}$ to ensure both low energy and minimal annihilation contribution. In our simulations, $\mathcal{N_Z} \sim 0.001$ for $m = 0.1, \varepsilon=1.0$, and $\sim 0.01$ for $m = 0.1, \varepsilon=0.2$. For $k \neq 0$, states with momentum $k$ and $-k$ have the same energy and can be distinguished by the eigenvalue of the charge conjugation operator: $c e^{-ik}$.

\begin{figure}[htp!]
    \centering
    \includegraphics[width=1\linewidth]{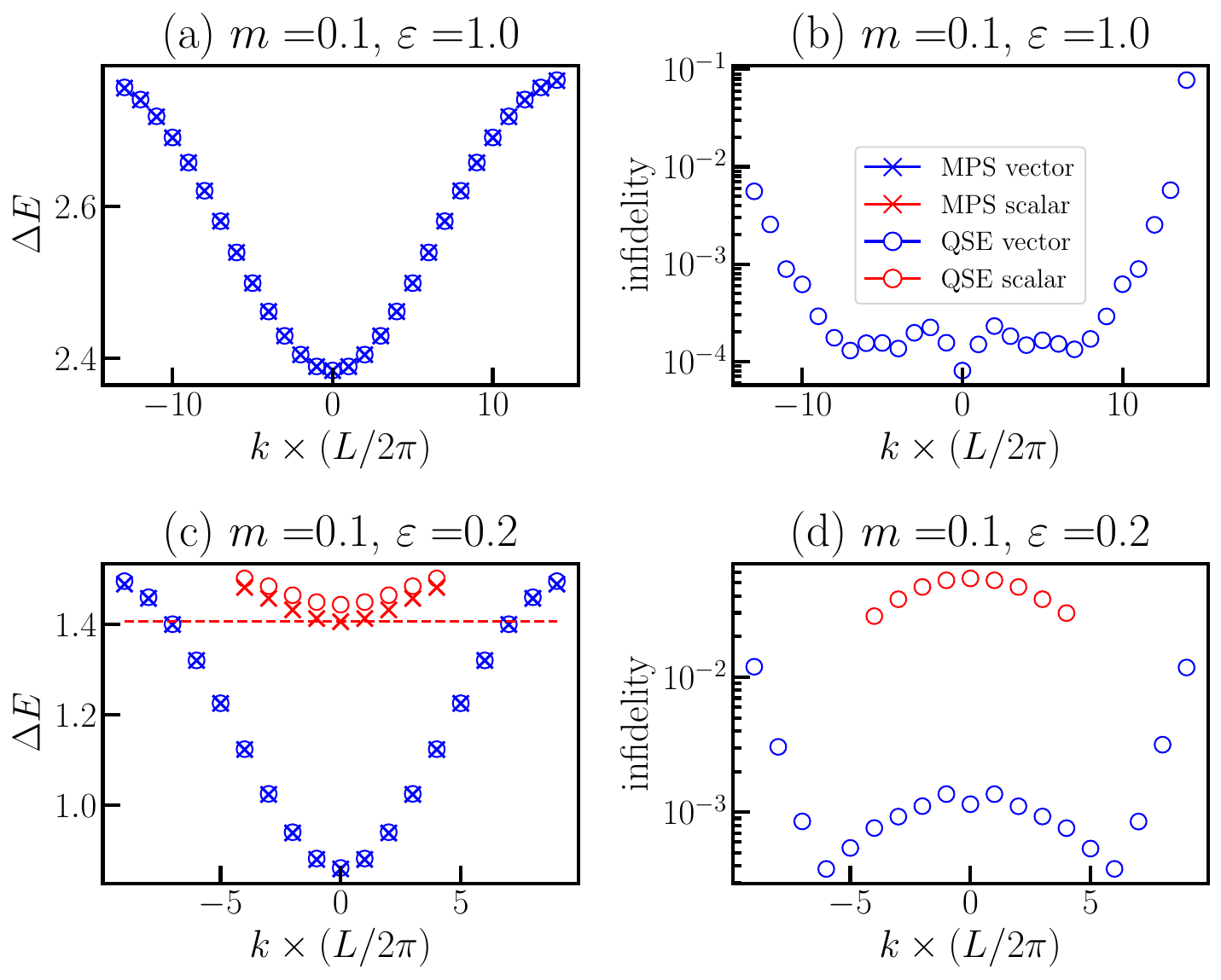}
    \caption{Energy difference of the excited states with respect to the ground state for $L=30$, $m=0.1$ and $\varepsilon=1.0$ (a), and $0.2$ (c). The blue (red) markers correspond to the vector (scalar) meson excitations, circles indicate the energies obtained by QSE, and crosses indicate those from MPS simulations. The $x$-axis is the corresponding integer lattice momenta $k$ in units of $L/2\pi$. Panels (b) and (d) show the corresponding infidelity between the states obtained by our QSE approach and the excited states obtained directly from the MPS simulation. The red dashed horizontal line in panel (c) indicates the mass of the scalar meson obtained from the MPS approach.}
    \label{fig:meson_spectrum}
\end{figure}

In order to demonstrate our approach produces high-fidelity mesonic excitations, we use MPS and first compute the ground state of the Hamiltonian using standard variational methods~\cite{Fishman_2022}. To obtain the excited states, we proceed in two ways. First, we use again standard variational MPS methods to compute the excited states of the model. Second, we create the subspace from Eq.~\eqref{eq: QSE_basis} explicitly by applying the meson operators on the MPS ground state. This allows us to construct the matrices in Eq.~\eqref{eq: QSE} and to solve the generalized eigenvalue problem and to compare the results from our method to the MPS data. In Fig.~\ref{fig:meson_spectrum}, we show the energy difference $\Delta E$ with respect to the ground state and the infidelity for the lowest-lying 28 excited states obtained from both methods for a system with $L=30$, $m=0.1$ and various values of $\varepsilon$. Note that the data for $k=0$ corresponds to the meson's mass in units of the lattice spacing. Focusing on strong coupling, $\varepsilon=1.0$, Fig.~\ref{fig:meson_spectrum}(a) shows that the first 28 excitations are vector mesons, and, as expected, the ones with $k\neq 0$ appear in pairs having the same absolute value of the momentum due to translation invariance. The energies obtained from our QSE method are in excellent agreement with the MPS solution. Looking at infidelities in Fig.~\ref{fig:meson_spectrum}(b) we observe that the meson excitations constructed using our subspace expansion approach are close to the MPS solution and for the low-lying states we obtain fidelities exceeding or approaching $99.9\%$. Only for excitations with large absolute momentum the accuracy drops slightly and for almost all states we have infidelities smaller than $1\%$. 

At smaller values of the coupling, $\varepsilon=0.2$, we see a similar vector meson branch as for the larger coupling, but the meson mass is significantly lighter than in the previous case. In addition, also scalar particles appear within the first 28 excited states (see Fig.~\ref{fig:meson_spectrum}(c)). The fidelities for the vector meson excitations obtained from QSE show a similar behavior as for the larger value of the coupling and for most of it above $99\%$. For the scalar mesons, the fidelity is slightly worse than for the vector branch, but still does not drop below $94.6\%$ despite us choosing the simplest mesonic operators in the ansatz in Eq.~\eqref{eq: bdag}. Using more elaborate operators in the ansatz, the fidelity for the scalar mesons could potentially also be increased to a similar level as for the vector branch.

\begin{figure}[htp!]
    \centering
    \includegraphics[width=1.0\linewidth]{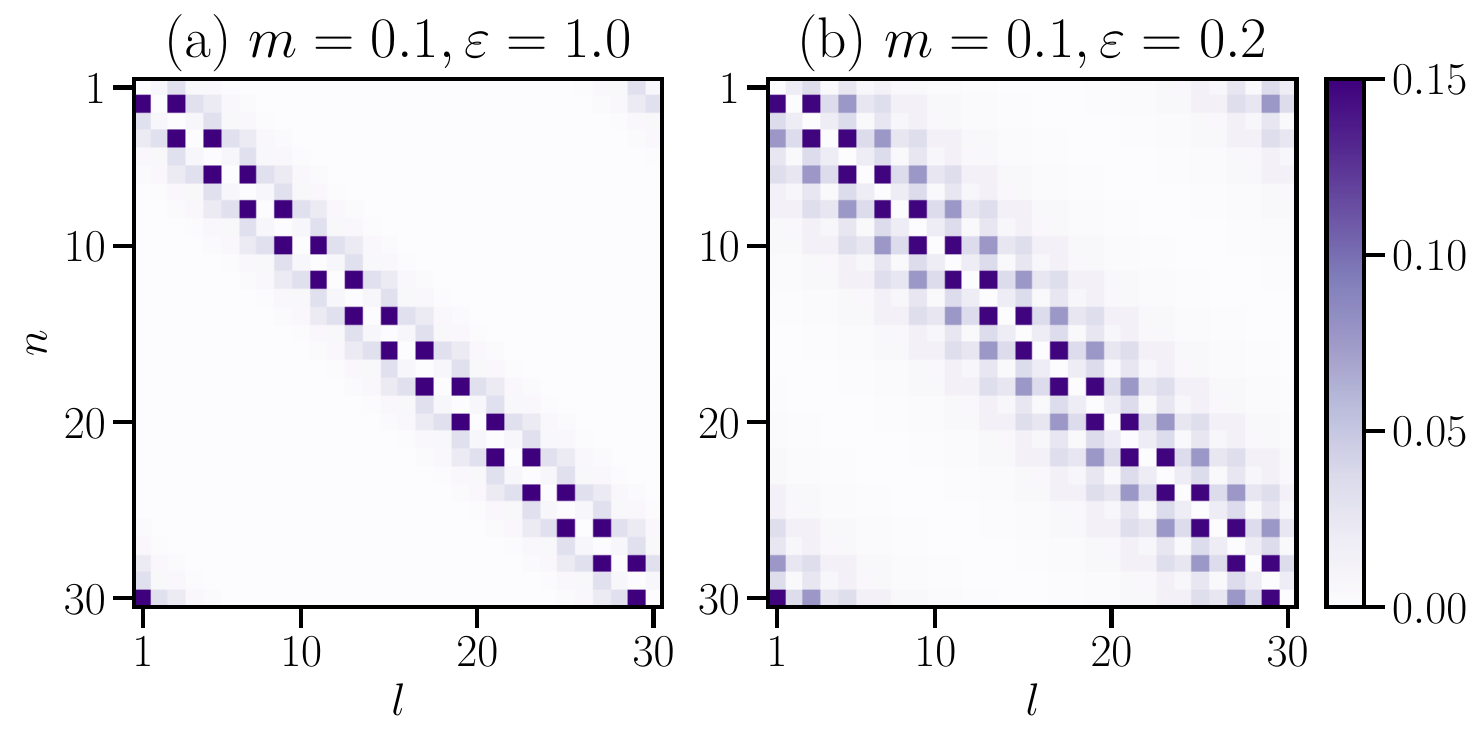}
    \caption{Distribution of vector meson's coefficients for $k=0$ at (a) $\varepsilon=1.0$ and (b) $\varepsilon=0.2$. The color bar represents the absolute value of coefficients, $|a^{0, -1}_{n,l}|$.}
    \label{fig: meson_coeffs}
\end{figure}

To explore the vector meson's structure at different values of the coupling, we present the distribution of the coefficients $a_{n,l}^{0,-1}$ obtained from Eq.~\eqref{eq: QSE} in Fig.~\ref{fig: meson_coeffs}. Looking at the vector meson at $k=0$ for a large coupling, $\varepsilon = 1.0$, Fig.~\ref{fig: meson_coeffs}(a) reveals that the coefficients for operators acting the nearest-neighbor matter sites dominate, i.e., $a_{n, n-1}^{0, -1}$ and $a_{n, n+1}^{0, -1}$. This indicates the meson is primarily composed of fermion-antifermion pairs connected by one excited electric link, as one might expect from the strong coupling picture in Fig.~\ref{fig:Z2LGT}. In contrast, at a weaker coupling of $\varepsilon = 0.2$ the vector meson exhibits a looser structure, allowing the fermion-antifermion pairs to separate further and be connected by longer electric links. This results in more coefficients having significant values, as Fig.~\ref{fig: meson_coeffs}(b) shows.
    
\section{Meson scattering}\label{sec: scattering}
Using the meson operators constructed in the previous section, we can now build operators creating vector meson wave packets. To this end, we define
\begin{equation}\label{eq: B_op}
    \begin{aligned}
        B^{\dagger}_{\bar{k}, \bar{x}} &= \sum_{k \in \Lambda^{\ast}} \phi(k)_{\bar{k}, \bar{x}}  b_{k, -1}^{\dagger}, \\
        &= \sum_{k \in \Lambda^{\ast}} \sum_{nl} \phi(k)_{\bar{k}, \bar{x}}  a^{k,-1}_{nl}\xi_n^{\dagger} W_{n, l} \xi_l,\\
        \phi_{\bar{k}, \bar{x}}(k) &= \frac{1}{\sqrt{\mathcal{N}_{\phi}}} e^{-ik\bar{x}} e^{-(k-\bar{k})^2/(4\sigma_k^2)},
    \end{aligned}
\end{equation}
where $ \phi_{\bar{k}, \bar{x}}(k)$ is a complex Gaussian distribution, $\bar{x}$, $\bar{k}$ represent the mean values of the wave packet's position and momentum, and $\sigma_k$ corresponds to its width in momentum space. Moreover, $\mathcal{N}_{\phi} = \sum_k |\phi(k)|^2$ is the normalization factor of the Gaussian distribution.
Note that $\Lambda^{\ast}$ does not correspond to the entire momentum space but rather to the momenta $k \in \Lambda$ that appear when solving Eq.~\eqref{eq: QSE}. For the example for $\varepsilon = 0.2$ shown in Fig.~\ref{fig:meson_spectrum}(c), $\Lambda^{\ast} = (2\pi/L) \times \{-9, -8, \cdots , 9\}$ for the vector meson, and $\Lambda^{\ast} = (2\pi/L) \times \{-4, -3, \cdots , 4\}$ for scalar meson. Similarly, for the example with $\varepsilon = 1.0$, for which only the vector meson appears (c.f.\ Fig.~\ref{fig:meson_spectrum}(a)), one obtains $\Lambda^{\ast} = (2\pi/L)\times \{-13, -12, \cdots , 14\}$. The restriction of the momentum modes in Eq.~\eqref{eq: B_op} will result in a wave packet that is not perfectly Gaussian. 
Due to the wave packet's localization in momentum space, the truncation error introduced by a finite momentum basis $\Lambda^{\ast}$ remains small as long as the central momentum $\bar{k}$ is far away from boundaries of $\Lambda^{\ast}$. In such cases, momenta near the cutoff are naturally suppressed by the Gaussian distribution, and the resulting wave packet remains well defined. However, for wave packets with higher momenta, e.g., $|\bar{k}| = (2\pi/L) \times 8$ with $m=0.1, \varepsilon=0.2$, the momentum cutoff will introduce noticeable truncation effects. This leads to an effectively narrower wave packet in momentum space and a broader one in position space. These effects can be mitigated by including more excited states in the subspace or using larger system sizes, which offer finer momentum resolution for defining narrow wave packets and enough spatial extent to separate broad wave packets. While we do not pursue these directions here due to computational constraints, they offer promising avenues for future work aiming to study high-momentum scattering processes more accurately.

Using the operator defined in Eq.~\eqref{eq: B_op}, one can prepare an initial state with two vector mesons to study their scattering processes. More specifically, for time $t=0$, the initial state we consider is given by 
\begin{equation}\label{eq: initial_state}
    \ket{\psi(t = 0)} = B_{\bar{k}, \bar{x}_1}^{\dagger} B_{-\bar{k}, \bar{x}_2}^{\dagger} \ket{\Omega},
\end{equation}
where the two wave packets are initially localized at $\bar{x}_1$ and $\bar{x}_2$, with momentum of equal magnitude $\bar{k}$ but opposite signs and the identical width $\sigma_k$. The time evolution $\ket{\psi(t)} = e^{-iHt}\ket{\psi(0)}$ can be computed with standard methods, e.g.\ using Trotterization.

To demonstrate that our method enables the preparation of wave packets and the investigation of scattering processes, we examine the collision of two mesons. Specifically, we consider a system with $L = 30$, which corresponds to $N = 60$ qubits, and we choose $\bar{x}_1 = 8$ and $\bar{x}_2 = 23$ and a width $\sigma_k = 2\pi / L$. To compute the time evolution, we use time-evolution block decimation where we employ a second order Trotterization, as detailed in App.~\ref{app: MPS_detals}. In order to characterize the scattering process, we monitor various observables. In particular, we compute the site-resolved flux configuration, $\bra{\psi(t)} X_{g, n} \ket{\psi(t)}$, and measure the staggered fermion density operator $\chi_n $ on each site~\cite{Davoudi_2024}
\begin{equation}
    \chi_n = \begin{cases} 
        \displaystyle 1 - \xi_n^{\dagger} \xi_n,  &n \in \text{odd}, \\
        \displaystyle   \xi_n^{\dagger} \xi_n , & n \in \text{even}.
    \end{cases}
\end{equation}
To gain insight into the energy transfer during the scattering process, we track the time-dependent contributions of the kinetic, mass, and electric components of the Hamiltonian. More specifically, we monitor
\begin{equation}
    \delta E_{B}(t) = \bra{\psi(t)} H_{B} \ket{\psi(t)} - \bra{\psi(0)} H_{B} \ket{\psi(0)},
\end{equation}
where $H_B$ is one of the terms in Eq.~\eqref{eq: Hamiltonian}, and the initial value at $t=0$ is subtracted to clearly show the variation of each energy component. In addition, we also monitor the number of vector and scalar mesons throughout the collision by computing
\begin{equation}\label{eq: meson_number}
    \begin{aligned}
        \rho_{c} = \sum_{k \in \Lambda^{\ast}} \bra{\psi(t)} b_{k, c}^{\dagger} b_{k, c} \ket{\psi(t)},
    \end{aligned}
\end{equation}
where $c=-1$ (+1) corresponds to the vector (scalar) mesons. This allows us to draw conclusions if vector mesons are annihilated during the scattering process at the expense of creating scalar mesons. Furthermore, during the scattering process, two localized mesons can merge due to fermion-antifermion annihilation, resulting in an extended meson connected by a longer flux string. To probe this process and the subsequent string breaking~\cite{Alexandrou2025}, we calculate the probability $P_l$ of observing a single flux string of length $l$ at time $t$, given by
\begin{equation}\label{eq: P_l}
\begin{aligned}
    P_{l} &= \sum_{n=1}^{L-l} |\bra{\psi(t)} \xi_n^{\dagger} W_{n, n+l} \xi_{n+l} \ket{\Omega}|^2 \\
    &+ \sum_{n=l+1}^{L} |\bra{\psi(t)} \xi_n^{\dagger} W_{n, n-l} \xi_{n-l} \ket{\Omega}|^2
\end{aligned}
\end{equation}
where $l \in \{1, \cdots, L/2\}$ is restricted due to the periodic boundary conditions, consistent with Eq.~\eqref{eq: W_nl}. Note that we include both strings between fermion and antifermion (first line) as well as between antifermion and fermion (second line).

\begin{figure}[htp!]
    \centering
    \includegraphics[width=1.0\linewidth]{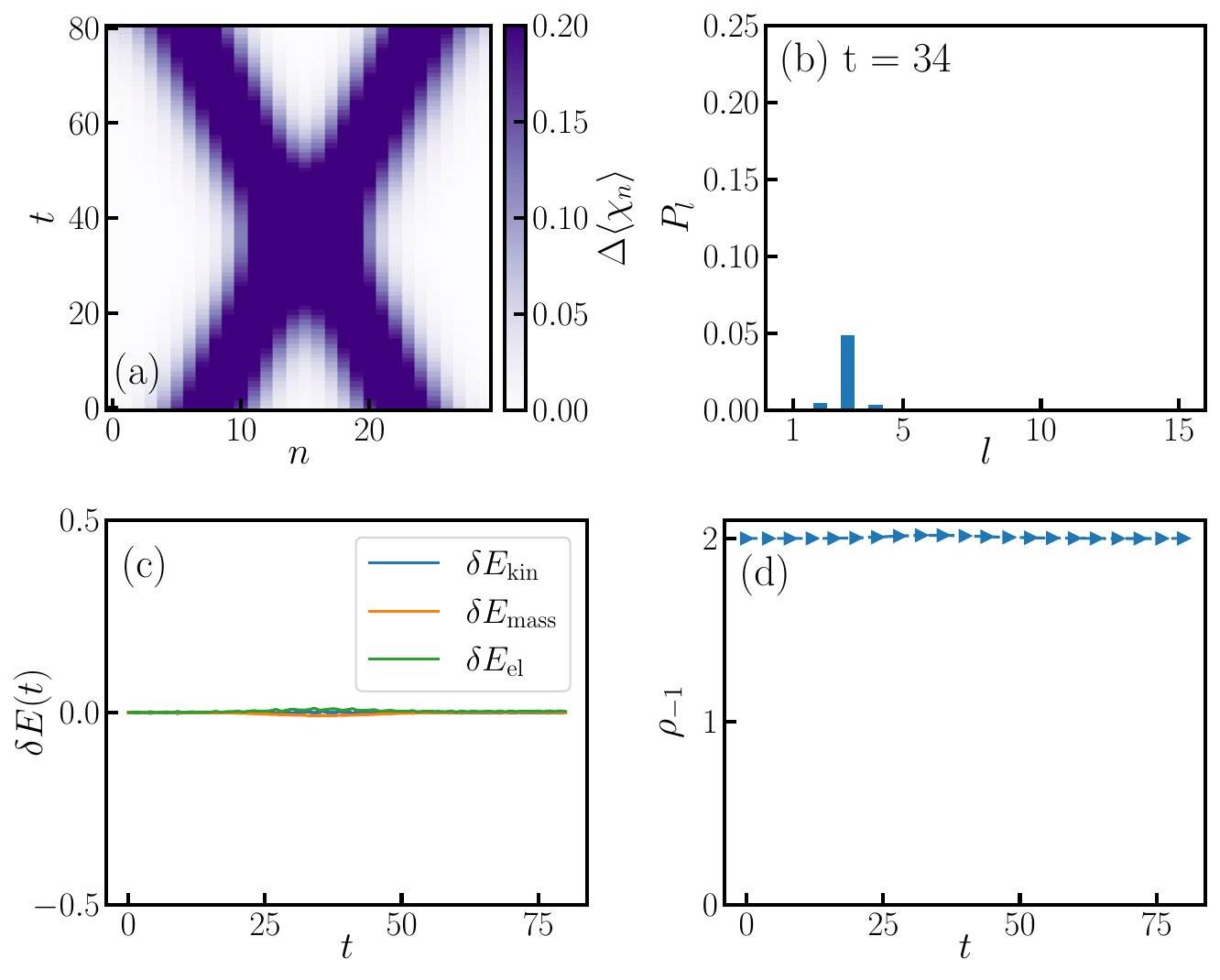}
    \caption{Time evolution of two meson wave packets with $m=0.1, \varepsilon = 1.0$, and momentum $\bar{k}_1 = 6 \times (2\pi/L)$. (a) Site-resolved staggered fermion density during the evolution. (b) Probability of having a single flux string with length $l$ at time $t=34$, which is around the collision of the two wave packets. (c) Change of kinetic, mass and electric field energies over time. (d) Vector meson number $\rho_{-1}$ over every four time steps. }
    \label{fig: elastic_scattering}
\end{figure}

First, we study the case of $m = 0.1$, $\varepsilon = 1.0$, for which we did not observe any scalar mesons in the low-lying spectrum (c.f.\ Fig.~\ref{fig:meson_spectrum}(a)). Figure~\ref{fig: elastic_scattering} displays our results for the evolution of two vector meson wave packets with opposite momenta  where $\bar{k} = 6 \times (2\pi/L)$. Panel~\ref{fig: elastic_scattering}(a) shows the site-resolved staggered fermion density over time, where we subtract the fermion density of the ground state and consider $\Delta\langle \chi_n\rangle = \langle\psi(t)|\chi_n|\psi(t) \rangle - \langle\Omega|\chi_n|\Omega \rangle$. As indicated by the fermion density, the two vector mesons propagate toward each other during the time evolution and eventually collide. The collision appears to be elastic, as there is no evidence of particle production in $\Delta\langle \chi_n\rangle$. This is further corroborated by the individual energy contributions in Fig.~\ref{fig: elastic_scattering}(c) and the vector meson number in Fig.~\ref{fig: elastic_scattering}(d). In particular, we observe that the individual parts of the Hamiltonian are almost conserved, and $\rho_{-1}$ has a value close to $2$ throughout the entire evolution, providing evidence that no particles are produced. In addition, we monitor the probability $P_l$ of forming a single flux string of length $l$ (as defined in Eq.~\eqref{eq: P_l}). At the beginning, the system consists of two separate flux strings coming from the two mesons, resulting in $P_l = 0$. As the two mesons approaching each other, these two strings merge to a single longer string, leading a peak in $P_l$ around $l=3$ at $t=34$ close to the collision, as Fig.~\ref{fig: elastic_scattering}(b) shows. After the collision, the flux string breaks, and $P_l$ returns to zero, indicating the hadronization to two separated mesons. The details of this string formation and breaking process can be found in Fig.~\ref{fig: strings_m0.1_eps_1.0_k6} in App.~\ref{app: string_details}.

\begin{figure*}
    \centering
    \includegraphics[width=0.95\linewidth]{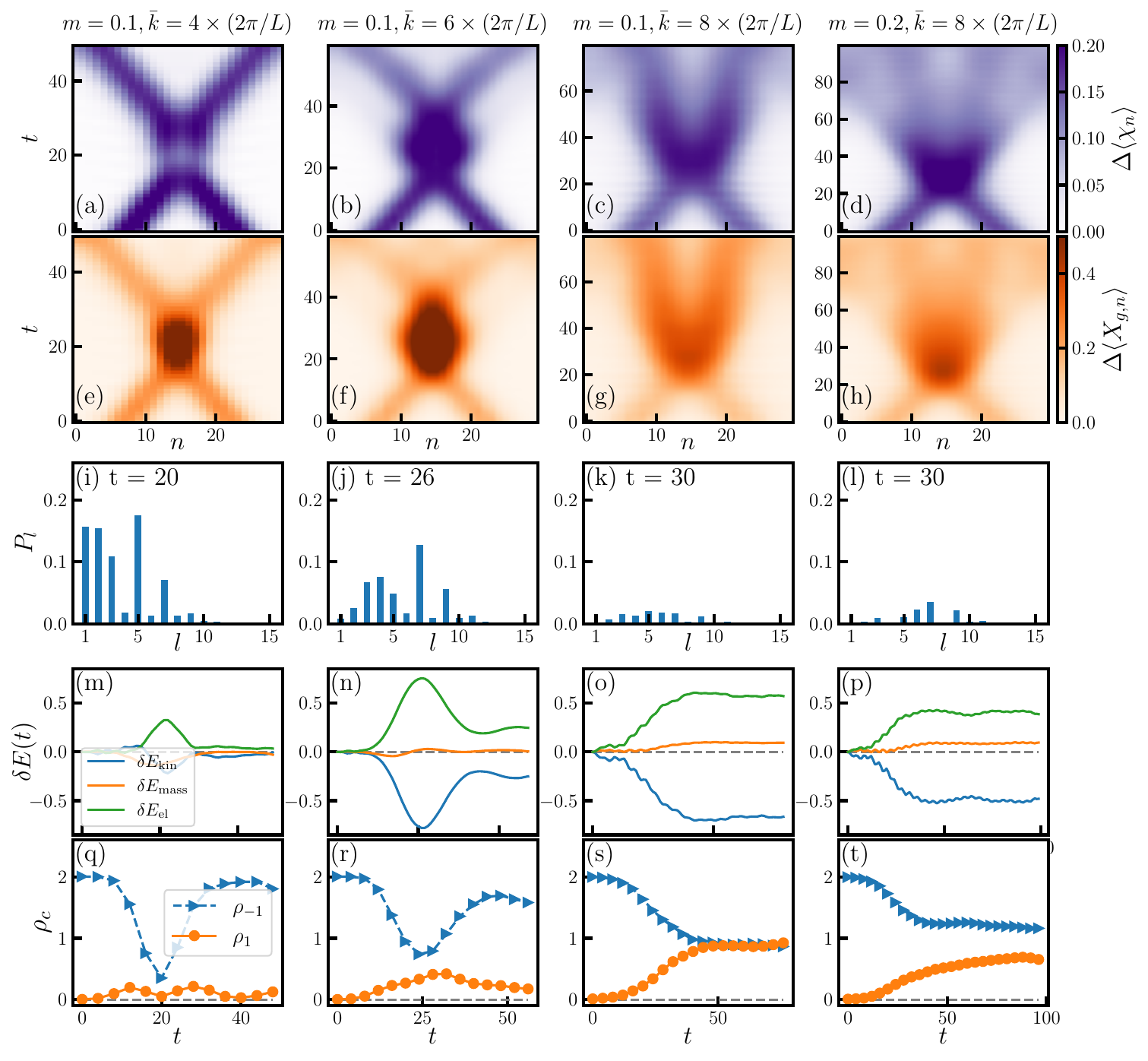}
    \caption{Scattering process between two vector meson wave packets with $\varepsilon=0.2$. Each column corresponds to different masses $m$ and initial momenta $\bar{k}$, as indicated in the title of each column. The first row shows the staggered fermion density; the second row shows the electric field density. Both are plotted as functions of position $n$ (x-axis) and time $t$ (y-axis), and share the same x-axis. The third row shows the probability of observing a single flux string at the collision point, plotted versus the string length $l$ (x-axis), with probability defined in Eq.~\eqref{eq: P_l}. The fourth row displays the time evolution of the kinetic, mass, and electric energy contributions, while the fifth row shows the number of vector and scalar mesons over time. The x-axis in the fourth and fifth rows is time $t$.}
    \label{fig: scattering_m0.1}
\end{figure*}

Second, we investigate the case of a smaller coupling, $m = 0.1$, $\varepsilon = 0.2$, for which we see a scalar meson in the low-energy spectrum (see Fig.~\ref{fig:meson_spectrum}(c)). Figure~\ref{fig: scattering_m0.1} shows the meson scattering process for various values of the momentum of the initial wave packets. In addition to the previous case, we also display the site-resolved electric field over time, where we again subtract the values of the ground state and consider $\Delta\langle X_{g,n}\rangle = \langle\psi(t)|X_{g,n}|\psi(t) \rangle - \langle\Omega|X_{g,n}|\Omega \rangle$. Focusing on the case of small momentum $\bar{k} = 4 \times (2\pi/L)$ first (c.f.\ left column of Fig.~\ref{fig: scattering_m0.1}), we observe indications for inelastic scattering. While there are no direct signs for the generation of new particles, the internal structure of the mesons changes during the collision. In particular, at the collision point around $t=20$, the staggered fermion density in Fig.~\ref{fig: scattering_m0.1}(a) shows a dip in the center of the system, which is accompanied by an increase in the electric field (c.f.\ Fig.~\ref{fig: scattering_m0.1}(e)). This indicates again fermion-antifermion annihilation and the formation of strings, which can be seen in Fig.~\ref{fig: scattering_m0.1}(i) showing noticeable probabilities for a single flux tube of length $l>1$. Compared to the case of larger coupling, longer strings are generated because of the smaller electric field energy contribution per link. After $t = 20$, the string breaks again and fermions are regenerated, which subsequently hadronize to two outgoing vector mesons (see Fig.~\ref{fig: strings_m0.1_eps_0.2_k4} in App.~\ref{app: string_details} for details). This observation is also confirmed by the different energy contributions in Fig.~\ref{fig: scattering_m0.1}(m) and the vector meson number in Fig.~\ref{fig: scattering_m0.1}(q). Around $t = 20$, both mass and kinetic energy decrease and are transformed into electric field energy. The formation of the string results in a significant reduction of the vector meson number, which is subsequently restored. Moreover, the scalar meson number $\rho_1$ stays close to 0 over the entire time and only shows a slight increase as the two vector mesons approach each other, thus confirming that essentially no scalar mesons are produced. The slight increase of $\rho_1$ after $t = 40$ can be attributed to the periodic boundary conditions, which cause the two vector mesons to approach each other again. 

Increasing the momentum to $\bar{k}= 6 \times (2\pi/L)$ (c.f.\ second column of Fig.~\ref{fig: scattering_m0.1}), we observe a fairly similar picture. However, due to the higher momentum, the resulting strings can extend to longer lengths, which can be observed in Fig.~\ref{fig: scattering_m0.1}(f) around $t = 20 \sim 30$ in form of an extended spatial region with high electric flux, and in Fig.~\ref{fig: scattering_m0.1}(j) showing dominating contribution for $l=7$. As shown in Fig.~\ref{fig: scattering_m0.1}(n), during the formation of the strings, the kinetic energy decreases and the electric field energy increases since longer flux tubes require higher energy. After the kinetic energy is exhausted, the fermion-antifermion pairs at the string's endpoints are pulled back towards each other and collide again. From $t = 40$ onwards, there remain some fermion-antifermion pairs connected with flux tubes in the middle of the system (see also Fig.~\ref{fig: strings_m0.1_eps_0.2_k6} in App.~\ref{app: string_details}), but most of the fermions hadronize to vector mesons going out. According to Fig.~\ref{fig: scattering_m0.1}(r), there is a signal that some scalar mesons are generated during the collision. However, $\rho_1$ decreases again significantly after $t=40$, indicating that no stable scalar mesons are formed.

Studying an even larger momentum of $\bar{k} = 8 \times (2\pi/L)$, one enters a regime where the vector meson has a higher energy than a scalar meson with $k=0$, as shown by the dashed line Fig.~\ref{fig:meson_spectrum}(c). Thus, we expect that stable scalar mesons can form during the scattering process in this case. Looking at the fermion density and the electric field in Figs.~\ref{fig: scattering_m0.1}(c) and \ref{fig: scattering_m0.1}(g), we indeed observe two outgoing particles with a slower velocity and stronger electric field after the collision. Contrary to the previous case, during the collision we do not observe a significant probability for a single flux string in Fig.~\ref{fig: scattering_m0.1}(k) and the electric field at the collision point is significantly smaller than before. Focusing on the different energy contributions in Fig.~\ref{fig: scattering_m0.1}(o), we see that the kinetic energy initially decreases and stabilizes at a value lower than the initial one after the collision, which is consistent with the smaller velocity of the outgoing particles. Most of this kinetic energy is transferred to the electric field energy, as the increase in $\delta E_\text{el}(t)$ shows. Besides, we observe a clear decrease of the vector meson number $\rho_{-1}$ from an initial value of 2 to 1 after the collision, where simultaneously the scalar meson number $\rho_1$ increases and also stabilizes around a value of 1 after the collision. This indicates that there are scalar mesons generated after the collision, while there is still a significant probability for two outgoing vector mesons. However, the staggered fermion density and the electric field density only display two fairly broad trajectories after the collision. This could be an effect of the resulting wave packets being spread out rather than remaining localized, and thus they overlap.

To obtain a clearer picture of the formation of scalar mesons, we repeat the simulation for  $\bar{k} = 8 \times (2\pi/L)$ but we use a slightly larger mass of $m=0.2$. This should result in the scalar mesons being heavier, thus having a smaller velocity after the collision and separating them better from the vector mesons. The results for this case are shown in the right column of Fig.~\ref{fig: scattering_m0.1}. While the results are qualitatively similar to that of the smaller mass $m=0.1$, the staggered fermion density and also the electric field now show indications for 4 outgoing trajectories (see panels (d) and (h)). Given energy conservation, these particles should consist of two heavier scalar mesons with low velocity and two vector mesons with higher velocity. While in this case the scalar meson number $\rho_1$ remains a little lower (see Fig.~\ref{fig: scattering_m0.1}(t)), the separation in velocity provides clearer evidence for the formation of stable scalar mesons in the scattering process.

\begin{figure}[htp!]
    \centering
    \includegraphics[width=0.95\linewidth]{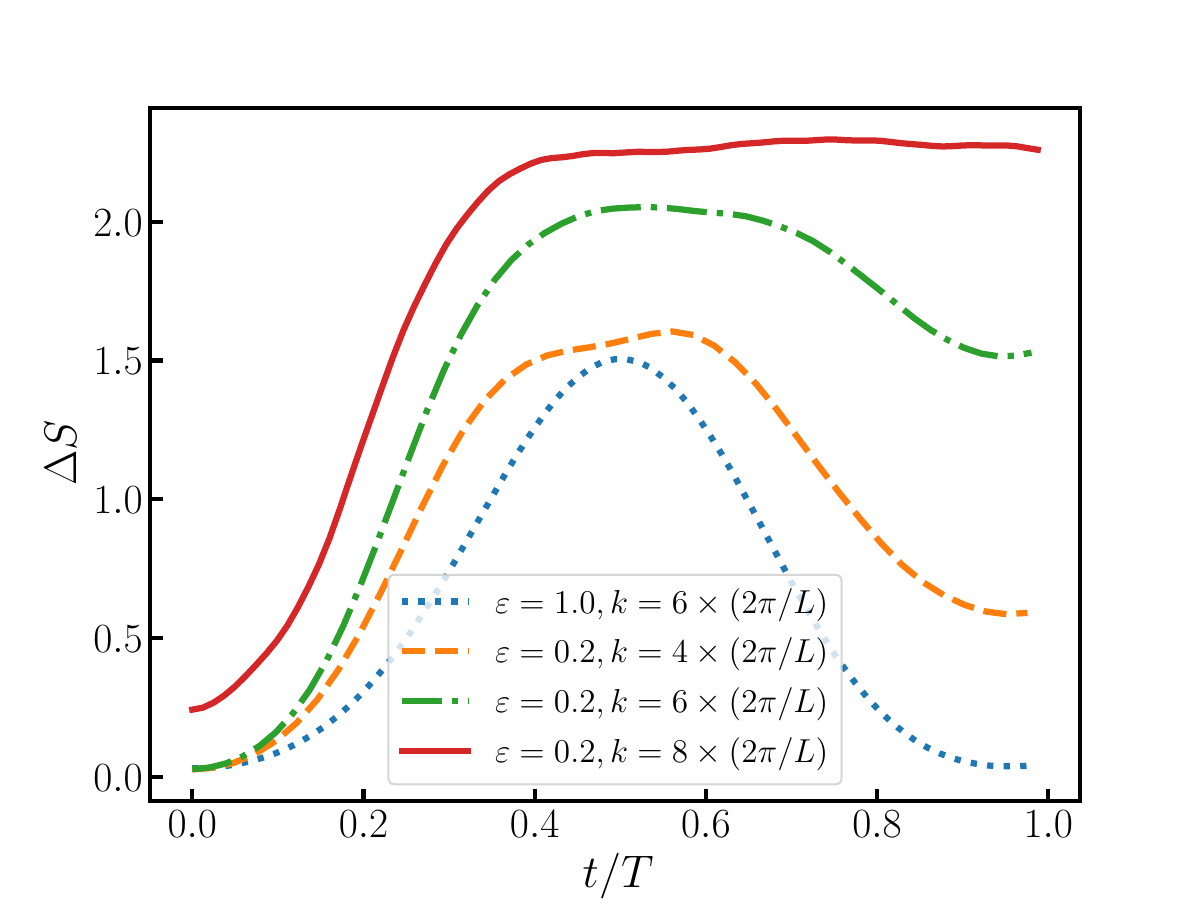}
    \caption{Bipartite entanglement entropy for both elastic and inelastic scattering processes with $m = 0.1$. Time is measured in units of the total evolution time $T$ to allow for comparing different scattering processes in a consistent manner. The final time $T$ is selected such that the particles have sufficient time to propagate after the collision, but short enough to prevent them from reaching the boundaries of the system. This choice is made to prevent an artificial increase in entanglement entropy that would result from the wave packets splitting as they cross the system's boundary.  The blue dotted line corresponds to the elastic scattering with $\varepsilon = 1.0, \bar{k}=6\times(2\pi/L)$, and $T = 70$.  The orange dashed, green dash-dotted and red solid lines are associated with the inelastic scattering with $\varepsilon = 0.2$, for momentum $\bar{k}=4\times(2\pi/L), 6\times(2\pi/L), 8\times(2\pi/L)$ and $T = 40$, $50$, and $80$  respectively.}
    \label{fig: entropy}
\end{figure}

Finally, we evaluate the bipartite entanglement entropy between the subsystems $\mathcal{L} = \{n <= L/2\}$ and $\mathcal{R} = \{n > L/2\}$. Specifically, we calculate the von Neumann entropy $S(t) = -\tr[ \tilde{\rho}(t) \log_2 \tilde{\rho}(t) ]$, where $\tilde{\rho}(t)$ is the reduced density matrix of one subsystem at time $t$, then subtract the vacuum contribution to obtain $\Delta S(t)$. We consider $m = 0.1$ and both elastic scattering with $\varepsilon = 1.0$ as in Fig.~\ref{fig: elastic_scattering}, and the inelastic scattering processes with $\varepsilon = 0.2$ as shown in the first three columns of Fig.~\ref{fig: scattering_m0.1}. For these processes, the evolution of the bipartition entropy is shown in Fig.~\ref{fig: entropy}. For the elastic case (blue dotted line), the entropy increases as the two mesons propagate toward the center and decreases after the collision, eventually resulting in nearly zero excess entropy relative to the vacuum. In contrast, for the inelastic scattering processes with $\varepsilon=0.2$, the peak in entropy at the collision point is larger and also the final value after the scattering process is higher, especially for a high momentum of $\bar{k} = 8 \times (2\pi/L)$. This indicates that the inelastic scattering will be more challenging for the numerical methods based on TNS. Specifically, in our MPS simulations we see that for reaching a given precision, the maximal size of the matrices required is around 55 for the elastic case (c.f.\ blue dotted line in Fig.~\ref{fig: entropy}), whereas during the simulation it reaches around $1000$ for the inelastic scattering associated with the red solid line in Fig.~\ref{fig: entropy}. 

\section{Preparing wave packets on a digital quantum computer}\label{sec: circ_wp}

In the previous sections, we discussed how to construct proper operators for creating meson wave packets and investigated scattering processes using MPS simulations. Our results show that inelastic scattering processes can be challenging to simulate classically with TNS, motivating the exploration of quantum computing as a promising alternative for these regimes. As outlined in Sec.~\ref{sec: scattering}, simulating the scattering process on a quantum computer requires first preparing the initial state comprising particle wave packets, followed by applying time evolution. The quantum circuit for time evolution can be implemented using Trotterization. However, constructing a quantum circuit for preparing wave packets is nontrivial, as the particle creation operator is not unitary. Reference~\cite{Davoudi_2024} has proposed a quantum circuit to prepare the meson wave packet approximately based on the Jordan, Lee, and Preskill (JLP) protocol~\cite{Jordan_2012, Jordan2019}, where the number of CNOT-gates scales as $O(L^3 N_{\mathrm{Trotter}})$, where $N_{\mathrm{Trotter}}$ is the number of Trotter steps. In this section, we propose an efficient approach to realize the meson wave packets with quantum circuit exactly, where the number of CNOT gates scale as $O(L^2)$ and the circuit depth is $O(L)$.

More specifically, the initial state we want to prepare consists of the vacuum state $\ket{\Omega}$ and meson creation operator $B^{\dagger}_{\bar{k}, \bar{x}}$ applied to it, as defined in Eq.~\eqref{eq: initial_state}. The vacuum state $\ket{\Omega}$ can be obtained by VQE~\cite{Peruzzo_2014, Farrell_2024} or other techniques~\cite{McClean2017,Yuan2019,Gacon2023}. The operator $B^{\dagger}_{\bar{k}, \bar{x}}$ is non-unitary, making it nontrivial to design a quantum circuit for its implementation. To address this problem, we propose decomposing these operators using Givens rotations~\cite{Wecker_2015, Kivlichan_2018, Jiang_2018}. For convenience, we define a hermitian operator $A_{\bar{k}, \bar{x}}^{\dagger} = B^{\dagger}_{\bar{k}, \bar{x}}  + B_{\bar{k}, \bar{x}}$. In principle, $A_{\bar{k}, \bar{x}}^{\dagger} \ket{\Omega} = B^{\dagger}_{\bar{k}, \bar{x}} \ket{\Omega}$, as $B_{\bar{k}, \bar{x}} \ket{\Omega} = 0$ due to the annihilation property $b_k \ket{\Omega} = 0$. Although this equality might be slightly violated when using the operator obtained via QSE, our numerical simulations show that the state $A_{\bar{k}, \bar{x}}^{\dagger} \ket{\Omega}$ achieves a fidelity of $99.9\%$ with the state $B_{\bar{k}, \bar{x}}^{\dagger} \ket{\Omega}$.

The hermitian operator $A_{\bar{k}, \bar{x}}$ can be expressed as
\begin{equation}
    A_{\bar{k}, \bar{x}}^\dagger =\sum_{nl} \mathcal{M}_{nl} \tilde{\xi}_n^{\dagger} \tilde{\xi_l},
\end{equation}
where $\mathcal{M}_{nl}$ are coefficients corresponding to the wave packet's amplitude $\phi_k^s$ and the coefficients of vector meson operators $a^{k, -1}_{nl}$, similar as in Eq.~\eqref{eq: B_op}. The fermionic operators $\tilde{\xi}_n^{\dagger}$, $\tilde{\xi}_n$ incorporate a string of link variables starting from site 1  
\begin{equation}\label{eq: gauged_fermion}
    \begin{aligned}
        \tilde{\xi}_n^{\dagger}  = \xi_n^{\dagger} W_{n, 1},\ \ \tilde{\xi}_n = W_{1, n} \xi_n. 
    \end{aligned}
\end{equation}
Thus, the operator $\tilde{\xi}_n^{\dagger} \tilde{\xi_l}$  is gauge invariant. The coefficient matrix $\mathcal{M}$ is hermitian and with dimension $L \times L$, hence it can be diagonalized as $\mathcal{M} = u D u^{\dagger}$, where $u$ is a unitary matrix and $D = \text{diag}(\lambda_1, \lambda_2, \cdots )$ is a real diagonal matrix. This allows us to rewrite the operator $A_{\bar{k}, \bar{x}}^{\dagger}$ as
\begin{equation}\label{eq: decomp_A}
    \begin{aligned}
        A_{\bar{k}, \bar{x}}^{\dagger} &= \sum_{nlr} \tilde{\xi}_n^{\dagger} (u_{nr} D_{rr} u^{\dagger}_{rl}) \tilde{\xi}_l,\\
        &= \sum_r \left( \sum_n \tilde{\xi}_n^{\dagger} u_{nr} \right) \lambda_r \left( \sum_l u^*_{lr} \tilde{\xi_l} \right),\\
        &= \sum_r \left( V(u, \tilde{\xi}) ~\tilde{\xi}_r^\dagger ~V(u, \tilde{\xi})^\dagger \right) 
        \lambda_r
        \left( V(u, \tilde{\xi}) ~\tilde{\xi}_r ~V(u, \tilde{\xi})^\dagger \right), \\
        &= V(u, \tilde{\xi}) O_D V(u, \tilde{\xi})^\dagger,
    \end{aligned}
\end{equation}
where $O_D=\left( \sum_r \lambda_r \tilde{\xi}_r^\dagger \tilde{\xi}_r \right)$ is diagonal in fermionic Fock basis for the matter sites and the Pauli-Z basis for the links. In the third line, we used the property that a linear combination of fermionic operators can be realized as a unitary transformation.
\begin{equation}\label{eq: linear_fermions}
    \begin{aligned}
        \sum_l \tilde{\xi}_l^{\dagger} u_{ln}  &= V(u, \tilde{\xi}) ~\tilde{\xi}_n^{\dagger} ~V(u, \tilde{\xi})^{\dagger}, \\
        \sum_l \tilde{\xi}_l u_{ln}^{\ast} &= V(u, \tilde{\xi}) ~\tilde{\xi}_n ~V(u, \tilde{\xi})^{\dagger},
    \end{aligned}
\end{equation}
with $V(u, \tilde{\xi})$ being an unitary operator defined as:
\begin{equation}\label{eq: V_u}
    V(u, \tilde{\xi}) =\exp\left(\sum_{nl}\tilde{\xi}^\dagger_n \left[\log u\right]_{nl} \tilde{\xi}_l\right).
\end{equation}
The proof of Eq.~\eqref{eq: linear_fermions} is provided at App.~\ref{app: proof_GivensRotations}. In the following, we will introduce the quantum circuit corresponding to Eq.~\eqref{eq: decomp_A}. Specifically, we will introduce the circuit for $V(u, \tilde{\xi})$ in subsection~\ref{subsec: circ_Vu}, and the circuit for diagonal part $O_D$ in subsection~\ref{subsec: circ_diag}. In the subsection~\ref{subsec: circ_whole}, we provide a resource estimation for the full circuit implementing the operator $A_{\bar{k}, \bar{x}}^{\dagger}$.

\subsection{Circuit for \texorpdfstring{$V(u, \tilde{\xi})$}{V(u, xi)}}\label{subsec: circ_Vu}
The operator $V(u, \tilde{\xi})$ satisfies the homomorphism property under matrix multiplication, i.e., $V(v, \tilde{\xi}) \times V(u, \tilde{\xi}) = V(v \times u, \tilde{\xi})$, which is proven in App.~\ref{app: proof_GivensRotations}. Utilizing this property,  $V(u, \tilde{\xi})$ can be decomposed into local operators with a QR-decomposition of $u$ via Givens rotations, as shown in Refs.~\cite{Kivlichan_2018, Jiang_2018}. In summary,  $V(u, \tilde{\xi})$ can be decomposed by $L (L-1) / 2$ local operators with $2L-3$ layers~\cite {Kivlichan_2018}. Specifically, the local operators corresponding to the Givens rotations are in the following form:
\begin{equation}\label{eq: Vr}
    \begin{aligned}
        &R_{nl}^{\dagger}(\theta, \tilde{\xi}) = \exp\left( \theta_{n,l} [\tilde{\xi}^{\dagger}_{n-1} \tilde{\xi}_n - \tilde{\xi}^{\dagger}_n \tilde{\xi}_{n-1}] \right)\\
                           &= \exp\left(\theta_{n,l} [\xi^{\dagger}_{n-1} Z_{g, n-1} \xi_n - \xi^{\dagger}_n Z_{g, n-1} \xi_{n-1}] \right),\\
                           &\overset{\mathrm{JWT}}{=} \exp(-i \frac{\theta_{n,l}}{2} [\sigma_{n-1}^x Z_{g, n-1} \sigma_n^x + \sigma_{n-1}^y Z_{g, n-1} \sigma_n^y]).
    \end{aligned}
\end{equation}
where $\theta_{n,l}$ is determined by the ratio of matrix elements in $u$, as detailed in App.~A of Ref.\cite{chai2024}. These operators are a specific instance of the general construction in Eq.\eqref{eq: V_u}, chosen to eliminate the matrix element $u_{nl}$ via Givens rotation\footnote{$R_{nl}^{\dagger}(\theta, \tilde{\xi})$ corresponds to $V^{\dagger}(r_{n,l})$ as Eq.~(44) in Ref.\cite{chai2024}}. For clarity, we omit the single-qubit rotation gates associated with the complex phases of the matrix elements, which are also described in App.~A of Ref.\cite{chai2024}. The final expression in Eq.\eqref{eq: Vr} follows from applying the Jordan-Wigner transformation in Eq.~\eqref{eq: JWT}.

\begin{figure}[htp!]
    \centering
    \includegraphics[width=1.0\linewidth]{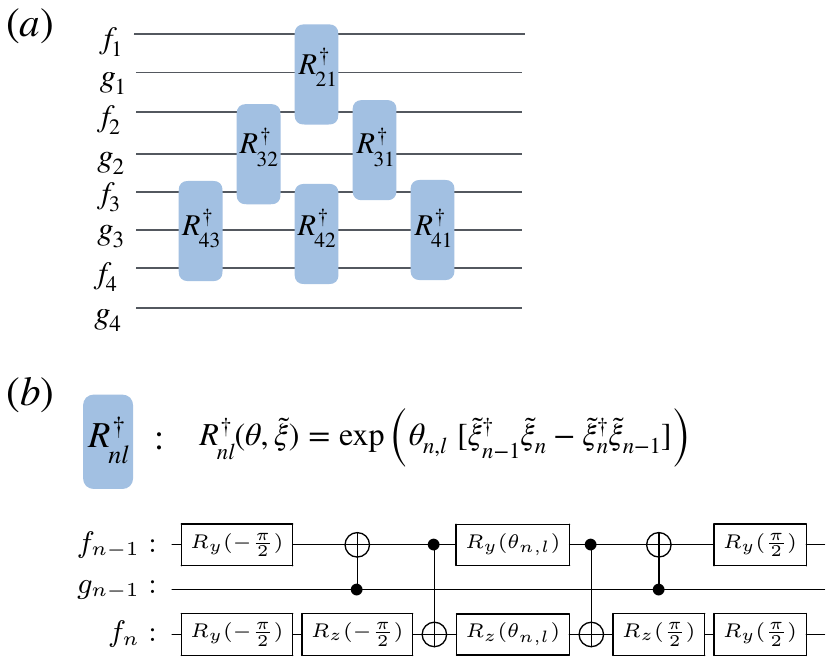}
    \caption{(a) Decomposition of $V(u, \tilde{\xi})$ for a system with $L = 4$ using six Givens rotations. (b) Decompose the local Givens rotations $R_{nl}^{\dagger}(\theta, \tilde{\xi})$ into 4 CNOT gates and single-qubit rotation gates. Qubits labeled as $f_n$ represent the matter field at site $n$, and the qubits labeled as $g_{n-1}$ represent the gauge field on the link between the sites $n-1$ and $n$.}
    \label{fig: Vu_circuit}
\end{figure}

To illustrate the result, we take an example with $4$ sites and show the circuit for $V(u, \tilde{\xi})$ in Fig.~\ref{fig: Vu_circuit}(a). The decomposition of  $R_{nl}^{\dagger}(\theta, \tilde{\xi})$ in standard single and two-qubit gates is depicted in Fig.~\ref{fig: Vu_circuit}(b), and comprises $4$ CNOT gates. In total, the decomposition of $V(u, \tilde{\xi})$ for a system with $L$ sites requires $2L(L-1)$ CNOT gates and results in a circuit depth of $8L-12$.

\subsection{Circuit for the diagonal operator $O_D$}\label{subsec: circ_diag}
Furthermore, the diagonal part $O_D$ in Eq.~\eqref{eq: decomp_A} also needs to be implemented with a quantum circuit. After applying the JWT, it reads
\begin{equation}
    \begin{aligned}
        O_D = \frac{I}{2}\sum_n \lambda_n + \frac{\lambda_1}{2} \sigma_1^z + \frac{\lambda_2}{2} \sigma_2^z + \cdots + \frac{\lambda_L}{2} \sigma_L^z.
    \end{aligned}
\end{equation}
To find a circuit implementing $O_D$, we consider a more general expression, which includes the above equation as a special case
\begin{equation}\label{eq: O_D}
    \begin{aligned}
        O_D  &= \sum_{n < L} \lambda_n P_n, 
    \end{aligned}
\end{equation}
with $ P_n \in \{I, \sigma^x, \sigma^y, \sigma^z \}^{\otimes L}$ a Pauli string and we require $ [P_n, P_l] = 0$. To realize such a linear combination of Pauli operators, we need to define some ancillary fermionic operators on the qubits representing the gauge field, which will ultimately have no effect on these qubits. To distinguish these from the fermionic operators $\xi$ representing the matter fields, we define these ancillary fermionic operators as
\begin{equation}
    \begin{aligned}
        \psi^{\dagger}_{g, n} &=  \prod_{l<n} \left( -iZ_{g, l}\right) \frac{X_{g,n} + iY_{g, n}}{2}, \\
        \psi_{g, n} &= \prod_{l<n} \left( iZ_{g, l}\right) \frac{X_{g, n} - iY_{g, n}}{2}, \\
        \tilde{X}_{g, n} &=\frac{\psi^{\dagger}_{g, n} + \psi_{g, n}}{\sqrt{2}}.
    \end{aligned}
    \label{eq:ancilla_fermions}
\end{equation}
where the operators $X_{g, n}, Y_{g, n}, Z_{g, n}$ are Pauli operators acting on the gauge field qubits as before. The operators defined in Eq.~\eqref{eq:ancilla_fermions} fulfill the anticommutation relations
\begin{equation}\label{eq: anticommute_X}
    \begin{aligned}
        \{\psi_{g, n}^{\dagger}, \psi_{g, l}\} = \delta_{n,l},\ \{\psi_{g, n}, \psi_{g, l}\} &= \{\psi_{g, n}^{\dagger}, \psi_{g, l}^{\dagger}\} =0, \\
        \{\tilde{X}_{g, n}, \tilde{X}_{g, l}\} &= \delta_{n,l}.
    \end{aligned}
\end{equation}
In addition, all of them commute with Pauli strings $P_n$ acting on the matter sites, as they operate on disjoint sets of qubits.

\begin{figure}[htp!]
    \centering
    \includegraphics[width=1.0\linewidth]{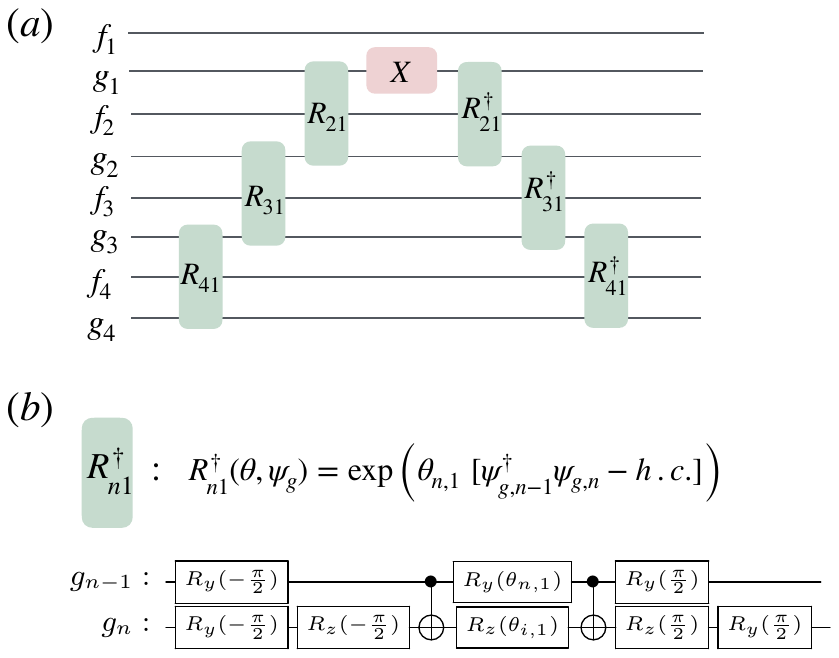}
    \caption{(a) Illustration of the circuit for $O_a$ for $L=4$. The first three Givens rotation gates correspond to the decomposition of $V(u_a, \psi_g)^{\dagger}$ and the last three are for $V(u_a, \psi_g)$. Again, the phase gates are not shown for simplicity. (b) Decomposition of a local Givens rotation gate $R_{n1}^{\dagger}(\theta, \psi_g)$, which performs the rotation between two nearest-neighbor qubits representing the gauge fields, and can be implemented with standard gates using $2$ CNOT operations.}
    \label{fig: circ_Oa}
\end{figure}

Using these, we can consider another set of operators given by
\begin{equation}\label{eq: O_ab}
    \begin{aligned}
        O_a &= \sum_n \operatorname{sgn}(\lambda_n) \sqrt{|\lambda_n|}\tilde{X}_{g, n},\\
        O_b &= \sum_n \sqrt{|\lambda_n|} P_n \tilde{X}_{g, n}
    \end{aligned}
\end{equation}
with real values $\lambda_n$. This is motivated by the fact that the values $\lambda_n$ in Eq.~\eqref{eq: decomp_A} correspond to eigenvalues of the hermitian matrix $\mathcal{M}$. In the expression above, $\operatorname{sgn}(\lambda_n)$ represent the sign of $\lambda_n$. These operators satisfy the identity
\begin{equation}
    \{ O_a, O_b \}= \sum_n \lambda_n P_n
\end{equation}
The above equation utilizes the property of Eq.~\eqref{eq: anticommute_X} and the final result is exactly $O_D$ in Eq.~\eqref{eq: O_D} as desired. Hence, the problem of realizing $O_D$ reduces to decomposing the operators $O_a$ and $O_b$ into a quantum circuit. Once these circuits are constructed, the above anticommutator can be realized using a Hadamard test. $O_a$ can be expressed as a linear combination of fermion operators and can be decomposed by Givens rotation, similar to Eq.~\eqref{eq: linear_fermions}
\begin{equation}
    \begin{aligned}
        O_a &= \sum_n \operatorname{sgn}(\lambda_n)\sqrt{|\lambda_n|} \tilde{X}_{g, n}, \\
            &= \sum_n \operatorname{sgn}(\lambda_n) \sqrt{ \frac{\lambda_n}{2} } \left(  \psi^{\dagger}_{g, n} + \psi_{g, n}  \right)\\
            &=  V(u_a, \psi_g) \left( ~\psi_{g,1}^{\dagger} + \psi_{g,1} \right) V(u_a, \psi_g)^{\dagger}, \\
            &=  V(u_a, \psi_g) X_{g, 1} V(u_a, \psi_g)^{\dagger},
    \end{aligned}
\end{equation}
where the first column of the matrix $u_a$ relates the coefficients in $O_a$, i.e., $1/\sqrt{2 \mathcal{N}_{\lambda}} \times \left(\operatorname{sgn}(\lambda_1) \sqrt{|\lambda_1|}, \operatorname{sgn}(\lambda_2)\sqrt{|\lambda_2|},\cdots \right)^T$, with $\mathcal{N}_{\lambda}$ being a normalization factor\footnote{For our purposes, the explicit form of $u_a$ is not of importance. We are concerned solely with the existence of a unitary matrix $u_a$ whose first column has the aforementioned entries. The remaining columns can be constructed to ensure that $u_a$ is unitary, for example, by applying the Gram-Schmidt orthonormalization process.}. Since we are only concerned with the first column of $u_a$, $V(u_a, \psi_g)$ can be decomposed with $L-1$ Givens rotations, consists of local operators as in Eq.~\eqref{eq: Vr}, but with operator $\psi_g$ in the formula, i.e.\ $R^{\dagger}_{n,1}(\theta, \psi_g)$. We show the circuit for $O_a$ in Fig.~\ref{fig: circ_Oa}(a), which consists of $2(L-1)$ Givens rotations, and can be decomposed into a total of $4(L-1)$ CNOT gates.

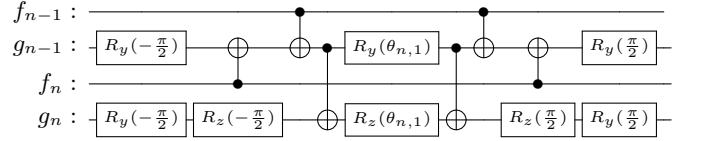
\begin{figure}[htp!]
    \centering
    \scalebox{1.0}{
    \Qcircuit @C=0.3em @R=0.1em @!R {
    \\
    \nghost{{q}_{0} :  } & \lstick{{f}_{n-1}:  } & \qw & \qw &\ctrl{1} & \qw &\qw &\qw &\ctrl{1} & \qw & \qw & \qw \\
    \nghost{{q}_{1} :  } & \lstick{ {g}_{n-1}: } & \gate{\scriptstyle{R_y}(-\frac{\pi}{2})} & \targ &\targ & \ctrl{2} & \gate{\scriptstyle{R_y}(\theta_{n,1})} & \ctrl{2} &\targ &\targ & \gate{\scriptstyle{R_y}(\frac{\pi}{2})}  & \qw & \qw \\
    \nghost{{q}_{2} :  } & \lstick{ {f}_{n}:  } & \qw & \ctrl{-1} & \qw & \qw & \qw &\qw &\qw &\ctrl{-1} & \qw & \qw & \qw \\
    \nghost{{q}_{3} :  } & \lstick{{g}_{n}  :  } & \gate{\scriptstyle{R_y}(-\frac{\pi}{2})} & \gate{\scriptstyle{R_z}(-\frac{\pi}{2})} &\qw &\targ & \gate{\scriptstyle{R_z}(\theta_{n,1})} & \targ &\qw & \gate{\scriptstyle{R_z}(\frac{\pi}{2})} & \gate{\scriptstyle{R_y}(\frac{\pi}{2})} & \qw & \qw \\
    \\
     }
     }
    \caption{Circuit of $R^{\dagger}_{n,1}(\theta, \tilde{\psi})$ for decomposing $O_b$, with operator $P_1 = I, P_n = \sigma^z_{n-1} $ for $n>1$ in Eq.~\eqref{eq: O_ab}.}
    \label{fig: circ_Ob}
\end{figure}

Regarding $O_b$, similar to the case of $\tilde{\xi}$, we can define the fermionic operator $\tilde{\psi}_{g, n} = P_n \psi_{g, n}$. Since $P_n^2 = I$, and $\tilde{\psi}_{g, n}, \tilde{\psi}_{g, n}^{\dagger}$ still satisfy the  standard fermionic anticommutation relations
\begin{equation}
    \begin{aligned}
        \{ \tilde{\psi}_{g, n}, \tilde{\psi}_{g, l}^{\dagger}\} &= P_n P_l \{ \psi_{g, n}, \psi_{g, l}^{\dagger}\} = \delta_{n,l}, \\
        \{ \tilde{\psi}_{g,n}, \tilde{\psi}_{g,l}\} &= P_n P_l \{ \psi_{g,n}, \psi_{g,l}\} = 0, \\
    \end{aligned}
\end{equation}
$O_b$ can be written as
\begin{equation}
    \begin{aligned}
        O_b &= \sum_n \sqrt{|\lambda_n|} P_n \tilde{X}_{g, n} \\
            &=  V(u_b, \tilde{\psi}_g) \left( \tilde{\psi}_{g,1}^{\dagger} + \tilde{\psi}_{g,1} \right) V(u_b, \tilde{\psi}_g)^{\dagger}, \\
            &=  V(u_b, \tilde{\psi}_g) X_{g, 1} V(u_b, \tilde{\psi}_g)^{\dagger}.
    \end{aligned}
\end{equation}
Here, the first column of $u_b$ is $1/\sqrt{2 \mathcal{N}_{\lambda}} \times \left(\sqrt{|\lambda_1|},\sqrt{|\lambda_2|},\cdots \right)^T$, and $P_1 = I$. Similar to $V(u_a, \psi)$, $V(u_b, \tilde{\psi})$ can be decomposed into $L-1$ Givens rotations, where each of these is implemented by a circuit like the one shown in Fig.~\ref{fig: circ_Oa}(a), but with the Givens rotation gates being $R^{\dagger}_{n,1}(\theta, \tilde{\psi})$
\begin{equation}
\begin{aligned}
    &R^{\dagger}_{n,1}(\theta, \tilde{\psi}) = \exp\left(\theta_{n,1} \left[ \tilde{\psi}^\dagger_{n-1}  \tilde{\psi}_{n} - \tilde{\psi}^\dagger_{n} \tilde{\psi}_{n-1} \right] \right),\\
    &= \exp\left( -i\frac{\theta_{n,1}}{2} P_{n-1} P_n \left[ X_{g, n-1} X_{g, n} + Y_{g, n-1} Y_{g, n} \right] \right).
\end{aligned}
\end{equation}
These can be implemented with $6$ CNOT gates as shown in Fig.~\ref{fig: circ_Ob}. All in all, $O_b$ can be implemented with $2(L-1)$ Givens rotation gates, and $12(L-1)$ CNOT gates.

\subsection{Full circuit for wave packet preparation}\label{subsec: circ_whole}
\begin{figure}[htp!]
    \centering
    \includegraphics[width=0.9\linewidth]{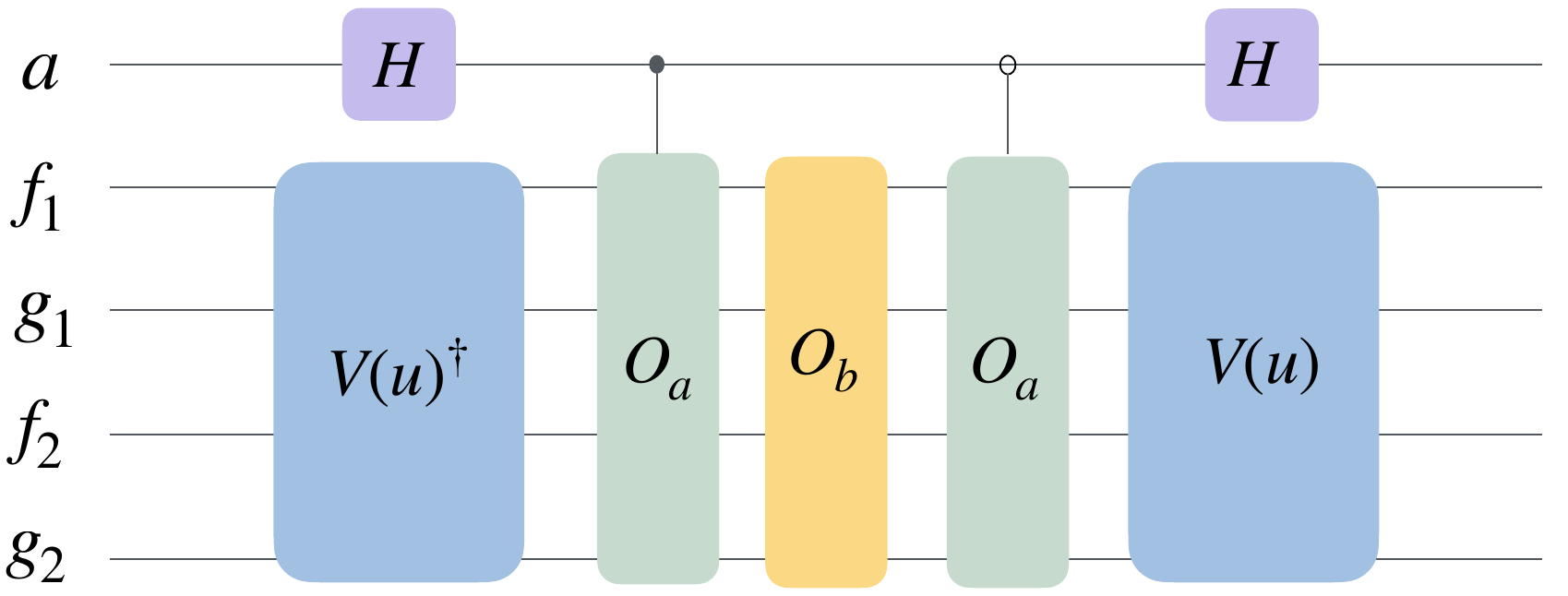}
    \caption{Sketch of the full circuit for preparing a meson wave packet. The qubit labeled $a$ is an ancilla qubit required for the Hadamard test. The individual operators $V(u)$, $O_a$, and $O_b$ can we decomposed in standard single and two-qubit gates as outlined in the text.}
    \label{fig: meson_circuit}
\end{figure}

Finally, we present the complete circuit for preparing a meson wave packet in Fig.~\ref{fig: meson_circuit}. 
Additionally, we summarize the number of CNOT gates and the circuit depths in Tab.~\ref{tab: gates_estimate}, assuming both nearest-neighbor and next-nearest-neighbor connectivity. It is worth noting that the estimation does not account for potential parallelization between different parts of the circuit, which could slightly reduce the circuit depth. Moreover, the depth could be further reduced with higher connectivity, which allows greater parallelization of Givens rotation. The anticommutator $\{O_a, O_b\}$ is implemented by the Hadamard test, where the control gate on $O_a$ only requires a controlled-$X$ gate, because the rest of the operator is identity if we remove the $X$ gate, as shown in Fig.~\ref{fig: circ_Oa}(a). Therefore, there are only $2$ extra CNOT gates introduced by the Hadamard test, we will ignore it in the overall resource estimation for simplicity.
\begin{table}[h!]
    \centering
        \begin{tabular}{@{} lcc @{}}
            \toprule
            & CNOT gates 
            & CNOT depth \\ \midrule
            $V(u,\tilde{\xi}) \text{ or } V(u,\tilde{\xi})^{\dagger}$ 
            & $2L(L-1)$ 
            & $4(2L - 3)$ \\ \midrule
            $O_a$ 
            & $4(L-1)$ 
            & $4(L-1)$ \\ \midrule
            $O_b$ 
            & $12(L-1)$ 
            & $12(L-1)$ \\ \midrule
            Total 
            & $4L^2 + 16L - 20$ 
            & $36L - 44$ \\ \bottomrule
    \end{tabular}
    \caption{Resource estimation for preparing a meson wave packet. The total two-qubit gate count and circuit depth include one $V(u,\tilde{\xi})$ and one $V(u,\tilde{\xi})^{\dagger}$, two $O_a$, and one $O_b$, as illustrated in Fig.~\ref{fig: meson_circuit}. For a narrow wave packet, $L$ can be replaced with $\Lambda \sim 1/\sigma_k$.}
    \label{tab: gates_estimate}
\end{table}

Note that for large systems and narrow wave packets, where $L \gg 1/\sigma_k$, only the qubits involved in the region in which the wave packet is nontrivial need to be considered in the circuit for preparing the wave packet. In such cases, the parameter $L$ in Table~\ref{tab: gates_estimate} can be effectively replaced by $\Lambda \sim 1/\sigma_k$.

\section{Summary and outlook\label{sec: summary}}
In this work, we introduce a comprehensive framework for the preparation of meson wave packets with well-defined momentum and charge conjugation number in a $\mathds{Z}_2$ lattice gauge theory. Utilizing a subspace expansion approach, we construct operators that allow for creating mesons on top of the ground state with high fidelity. By forming linear combinations of these operators with a Gaussian distribution, this technique facilitates the construction of spatially separated meson wave packets. Our approach is completely general and lends itself to any Hamiltonian-based numerical method. In particular, it can be directly used with TN and efficiently implemented on gate-based quantum computers using Givens rotation techniques, as we have shown.

Utilizing this technique, we construct two spatially separated meson wave packets with opposite momenta and simulate their real-time scattering dynamics using MPS. We observe both elastic and inelastic scattering processes across a range of model parameters and characterize the process in terms of local observables such as the site-resolved fermion and electric field density, as well as studying the energy transfer, meson number and the probability of observing a single flux string. These observables allow us to get a comprehensive insight into the scattering process. In particular, we observe the formation of strings during the collision followed by string breaking and hardronization. In the case of inelastic scattering, this process is accompanied by a significant increase in the electric energy contribution and a decrease in the kinetic energy, indicating the formation of new particles, which also show a clear signal in the meson number. Our results thus provide insight into the dynamics of inelastic scattering processes, inaccessible with conventional lattice methods. 

In addition, for the inelastic case we observe a significant growth in the bipartite entanglement entropy, which saturates at high values after the collision, thus motivating the implementation on quantum devices. To this end, we develop an efficient and accurate quantum circuit for meson wave packet preparation based on Givens rotations. For a system with $L$ sites, this decomposition yields a circuit with depth $\mathcal{O}(L)$ and requires $\mathcal{O}(L^2)$ CNOT gates, making it well-suited for implementations on near-term quantum devices. 

In this work, we used a $\mathds{Z}_2$ gauge theory as a representative example. The approach for generating meson wave packets is not limited to this setting and can be readily extended to other gauge models, such as the Schwinger model, and to higher-dimensional LGTs. Hence, it provides a stepping stone towards studying real-time dynamics of scattering processes in more complicated gauge theory, and for a novel characterization of complex phenomena, such as hadronization and fragmentation, beyond the capabilities of conventional lattice methods.

\paragraph*{Note added:}
Two related publications appeared at the same time as our work. Reference~\cite{schuhmacher2025} reports a quantum experiment of hadron scattering in a quantum link model using IBM's \texttt{ibm\_marrakesh} quantum computer. Reference~\cite{davoudi2025} presents a meson scattering experiment in the same $\mathds{Z}_2$ gauge theory as this work, performed on the trapped-ion platform \texttt{IonQ Forte}, with a focus on wave packet preparation and early-time dynamics. In contrast, our work develops a scalable framework for preparing particle wave packets, with improved methods for operator construction and circuit decomposition, and explores the full scattering process including inelastic dynamics and hadronization. Together with Ref.~\cite{davoudi2025}, our study represents a complementary effort, addressing both the algorithmic and experimental aspects of quantum simulations of scattering.\\

\begin{acknowledgments}
    The authors thank Lingxiao Xu and Karl Jansen for very helpful discussions.
    This work is supported with funds from the Ministry of Science, Research and Culture of the State of Brandenburg within the Center for Quantum Technology and Applications (CQTA). 
    \begin{center}
        \includegraphics[width = 0.08\textwidth]{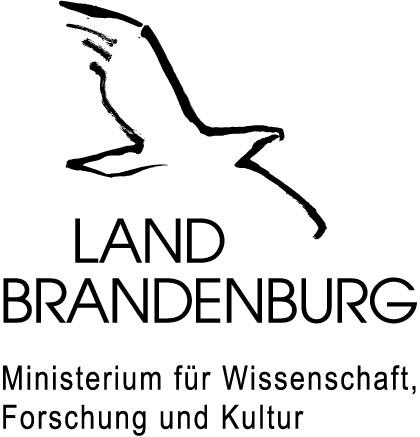}
    \end{center}
\end{acknowledgments}

\onecolumngrid
\appendix

\section{Quantum Subspace Expansion}\label{app: QSE}
In this appendix, we provide details on the QSE method used in our work. Reference~\cite{Yoshioka_2022_GQSE} offers a concise review of QSE in Appendix S1. Here, we extend the approach to include the additional constraints in our setup, namely the conditions that mesons should be an eigenstate of the charge conjugation operator $C$ and be annihilated by the operator $b_{k, c}$. As introduced in Ref.~\cite{Yoshioka_2022_GQSE}, a commonly employed strategy for obtaining the eigenstates of a Hamiltonian is based on the Ritz variational principle~\cite{Ritz1909}. This approach involves minimizing the following cost function for the quantum state $|\tilde{\psi}\rangle = \sum_I a_I M_I \ket{\Omega}$, defined within the subspace specified in Eq.~\eqref{eq: QSE_basis},
\begin{equation}\label{eq: lagrange}
    \begin{aligned}
        \mathcal{L} &= \langle \tilde{\psi} | H | \tilde{\psi} \rangle - \lambda \left( \langle \tilde{\psi} | \tilde{\psi} \rangle - 1 \right) \\
        &= \sum_{I, J} a_I^{\ast} a_J ~ \bra{\Omega} M_I^{\dagger}  H M_J \ket{\Omega} 
        - \lambda ~ \left( \sum_{I, J} a_I^{\ast} a_J  ~ \bra{\Omega} M_I^{\dagger}  M_J \ket{\Omega} - 1 \right),
    \end{aligned}
\end{equation}
where the $\lambda$ is the Lagrange multiplier to enforce the normalization of the wave function. From the stationarity condition, $\partial_{ a_{I}^{\ast} }\mathcal{L} = 0$, the following generalized eigenvalue equation can be derived
\begin{equation}
    \mathcal{H} \vec{a} = \lambda \mathcal{S} \vec{a},
\end{equation}
where $\mathcal{H}$ and $\mathcal{S}$ are matrix elements as in Eq.~\eqref{eq: QSE_HCSmat}. By solving the above equation, we can obtain the coefficients $\vec{a} = (a_{1,1}, a_{1,2}, \cdots, a_{2,1}, a_{2,2}, \cdots)$ that yield the eigenstate, with $\lambda$ being an eigenvalue $E$ of the Hamiltonian.

In this work, we also want impose that the solution is an eigenstate of the operator $C$, which commutes with Hamiltonian, hence there exists a common set of eigenfunctions. Besides, the annihilation condition, $b_{k, c} \ket{\Omega} = 0$, needs to be included, too. The corresponding generalization of Eq.~\eqref{eq: lagrange} then reads as 
\begin{equation}
    \begin{aligned}
        \mathcal{L} &= \bra{\Omega} b_{k,c} ~ H ~ b_{k,c}^{\dagger} \ket{\Omega} 
        + \bra{\Omega} b_{k,c} ~ C ~ b_{k,c}^{\dagger} \ket{\Omega} 
        + \langle \Omega|b_{k,c}^{\dagger} b_{k,c} | \Omega \rangle
        - \lambda \left( \bra{\Omega} b_{k,c} b_{k,c}^{\dagger} \ket{\Omega} - 1 \right) 
        \\
        &= \sum_{I, J} a_{I }^{(k,c) \ast} ~a_{J}^{(k,c)} ~ \bra{\Omega} M_I^{\dagger} H M_J \ket{\Omega} \\
        &+ \sum_{I, J} a_{I }^{(k,c) \ast} ~a_{J}^{(k,c)} ~ \bra{\Omega} M_I^{\dagger} C M_J \ket{\Omega} \\
        &+ \sum_{I, J} a_{I }^{(k,c) \ast} ~a_{J}^{(k,c)}  ~ \bra{\Omega} M_J  M_I^{\dagger} \ket{\Omega}\\
        &- \lambda ~ \left( \sum_{I, J} a_{I }^{(k,c) \ast} ~a_{J}^{(k,c)}  ~ \bra{\Omega} M_I^{\dagger} M_J \ket{\Omega} - 1 \right).
    \end{aligned}
\end{equation}
Here we label the coefficients with superscripts since the solution will also yield an eigenstate of $C$ with eigenvalue $c$ and momentum $k$. By imposing $\partial_{ a_{I }^{k,c \ast} }\mathcal{L} = 0$, we can obtain the following equation
\begin{equation}
    \left( \mathcal{H} + \mathcal{C} + \mathcal{Z} \right) ~\vec{a}^{(k,c)} = \lambda ~\mathcal{S} \vec{a}^{(k,c)},
\end{equation}
where the matrices $\mathcal{C}$, $\mathcal{Z}$ are the same as Eq.~\eqref{eq: QSE_Zmat}. The values $\lambda$ now include various contributions: $ \lambda = E + ce^{-ika} + \langle \Omega|b_{k,c}^{\dagger} b_{k,c} | \Omega \rangle$. Ideally, the last term should vanish, however, due to the finite basis employed, it generally yields nonzero, albeit small, values.

\section{Givens rotation for fermionic operators}\label{app: proof_GivensRotations}
References~\cite{Kivlichan_2018, Jiang_2018} have already demonstrated that the unitary transformation of the single-particle basis for fermions can be realized using Givens rotations. Here, we extend this idea from fermions to general fermionic operators, showing that Givens rotation can be applied to arbitrary operators, e.g.\ to fermion operators coupled with bosonic operators, as long as the anticommutation relations hold. Specifically, this applies to the operators $\tilde{\xi}_n$ defined in Eq.~\eqref{eq: gauged_fermion}. The proof of Eq.~\eqref{eq: linear_fermions} follows a similar approach to the appendix of Ref.~\cite{Kivlichan_2018}. For completeness, we present the proof below.

For the fermionic operators $\tilde{\xi}^{\dagger}_n, \tilde{\xi}_n$, the usual anticommutation relations are satisfied
\begin{equation}
    \{\tilde{\xi}^{\dagger}_n, \tilde{\xi}_l\} = \delta_{nl}, \ \ \{\tilde{\xi}_n, \tilde{\xi}_l\} = \{\tilde{\xi}^{\dagger}_n, \tilde{\xi}^{\dagger}_l\} = 0.
\end{equation}
Hence, for the operator $K = \sum_{nl} \log(u)_{nl} \tilde{\xi}_n^{\dagger} \tilde{\xi}_l$, with $u$ being a unitary matrix, the following equations hold
\begin{equation}
    [K, \tilde{\xi}^{\dagger}_r] = \sum_n  \tilde{\xi}^{\dagger}_n \log(u)_{nr}, \quad [K, \tilde{\xi}_r] = \sum_n \tilde{\xi}_n \log(u)^{\ast}_{nr}.
\end{equation}
Considering the transformation by operator $V(u, \tilde{\xi}) = e^{K}$ one finds
\begin{align}
    &V(u, \tilde{\xi})\tilde{\xi}_r^{\dagger} V(u, \tilde{\xi})^{\dagger} = e^{K} \tilde{\xi}_r^{\dagger} e^{-K} \\
    &= \tilde{\xi}_r^{\dagger} + [K, \tilde{\xi}_r^{\dagger}] + \frac{1}{2}[K, [K, \tilde{\xi}_r^{\dagger}]] \cdots \\
    &= \tilde{\xi}_r^{\dagger} + \sum_{n} \tilde{\xi}_n^{\dagger} \log(u)_{nr} + \frac{1}{2} \sum_{nl} \tilde{\xi}_l^{\dagger} \log(u)_{ln} \log(u)_{nr} \cdots \\
    &=\sum_n \tilde{\xi}_n^{\dagger} \{ \delta_{nr} + \log(u)_{nr} + \frac{1}{2}\log(u)^2_{nr} \cdots \} \\
    &=\sum_n \tilde{\xi}_n^{\dagger} u_{nr},
\end{align}
where we used the Baker-Campbell-Hausdorff expansion in the second line.

Next, we show that $V(u, \tilde{\xi})$ satisfies the homomorphism property under matrix multiplication. With two unitary matrix $u$ and $v$, $V(v, \tilde{\xi})\times V(u, \tilde{\xi}) = V(v\times u, \tilde{\xi})$, which can be proven as follows
\begin{equation}
    \begin{aligned}
     V(v, \tilde{\xi}) V(u, \tilde{\xi}) \tilde{\xi}_r^{\dagger} V(u, \tilde{\xi})^{\dagger} V(v, \tilde{\xi})^{\dagger} 
        &= V(v, \tilde{\xi}) \left( \sum_n \tilde{\xi}_n^{\dagger} u_{nr} \right) V(v, \tilde{\xi}) \\
        &= \sum_{nl} \tilde{\xi}_l^{\dagger} v_{ln} u_{nr}\\
        &= \sum_{l} \tilde{\xi}_l^{\dagger} \left( v u \right)_{lr}\\
        &= V(v u, \tilde{\xi}) \tilde{\xi}_r^{\dagger} V(v u, \tilde{\xi})^{\dagger}.
    \end{aligned}
\end{equation}
The above equation holds for arbitrary $\tilde{\xi}_r^{\dagger}$, so we conclude $V(v, \tilde{\xi}) \times V(u, \tilde{\xi}) = V(v \times u, \tilde{\xi})$.

\section{Single flux tube generation and string breaking\label{app: string_details}}
In this appendix, we present the probabilities $P_l$ for having a single flux tube of length $l$ during the scattering dynamics. To characterize the string generation and breaking, we select four representative time slices for all cases:  one before the collision, one during the collision, one shortly after the collision, and a final one the mesons have propagated for some time after the collision. The initial value of $P_l$ at $t=0$ is zero for all cases, because the initial state consists of two separate flux strings. Hence the probability of finding a single flux tube is zero, thus, we do not show these plots.
\begin{figure}[H]
    \centering
    \includegraphics[width=0.9\linewidth]{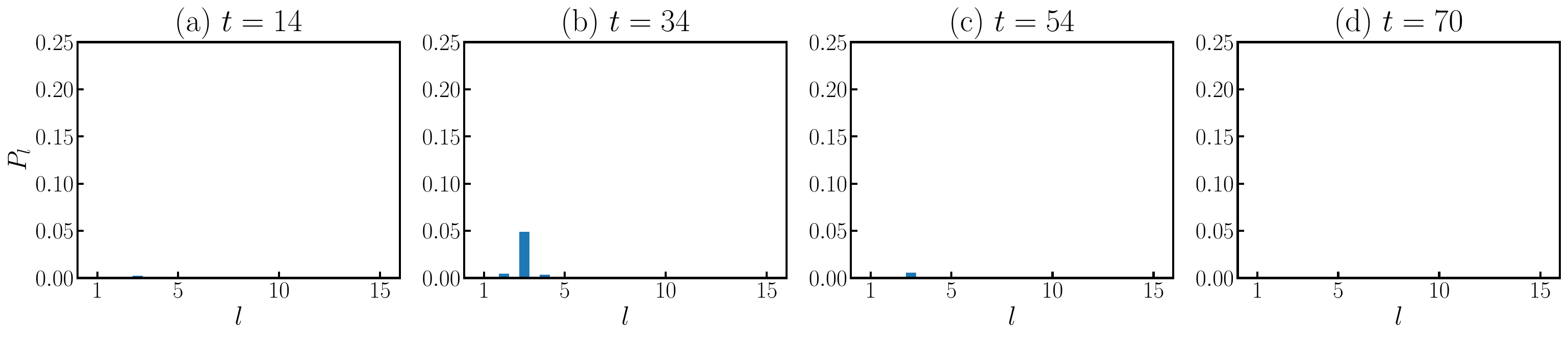}
    \caption{$P_l$ at different times with $m = 0.1, \varepsilon = 1.0, \bar{k} = 6\times(2\pi)/L$, corresponding to the elastic scattering process shown in Fig.~\ref{fig: elastic_scattering}. (a) Time slice before collision, where the two meson wave packets start to overlap, and the two separate flux strings associated with each meson start to merge into a longer single string. (b) Time slice during the collision, the overlap between two wave packets is enhanced and single flux tube is generated. (c) Time shortly after the collision, the single flux string breaks and corresponding probability decreases. (d) The time when two mesons are well separated, where the probability for a single flux string remaining is essentially zero, indicating that the strings broke and hadronized into the outgoing mesons.}
    \label{fig: strings_m0.1_eps_1.0_k6}
\end{figure}
\begin{figure}[H]
    \centering
    \includegraphics[width=0.9\linewidth]{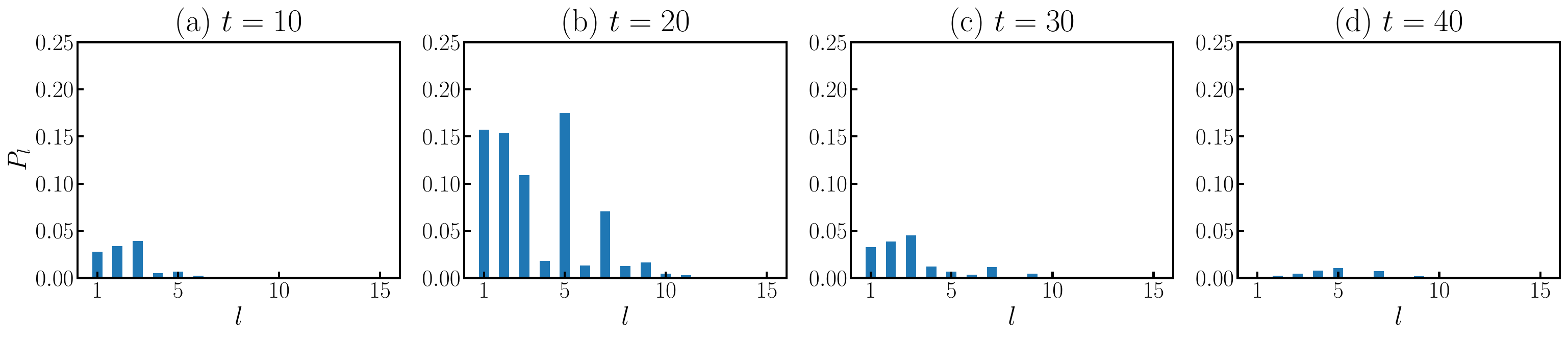}
    \caption{$P_l$ at different times with $m = 0.1, \varepsilon = 0.2, \bar{k} = 4\times(2\pi)/L$, corresponding to the inelastic scattering process shown in the first column of Fig.~\ref{fig: scattering_m0.1}. Compared to the elastic case, more and longer single flux strings are generated due to the smaller electric field energy. (a) Before the collision, the two mesons approach and longer strings begin to form. (b) During the collision, fermion-antifermion annihilation is enhanced as shown in Fig.~\ref{fig: scattering_m0.1}(a), leading to significant generation of strings. (c) Shortly after the collision, the single flux string breaks and the corresponding probabilities decrease. (d) When the two mesons are well separated, most strings are broken, indicating again hadronization into outgoing mesons.}
    \label{fig: strings_m0.1_eps_0.2_k4}
\end{figure}
\begin{figure}[H]
    \centering
    \includegraphics[width=0.9\linewidth]{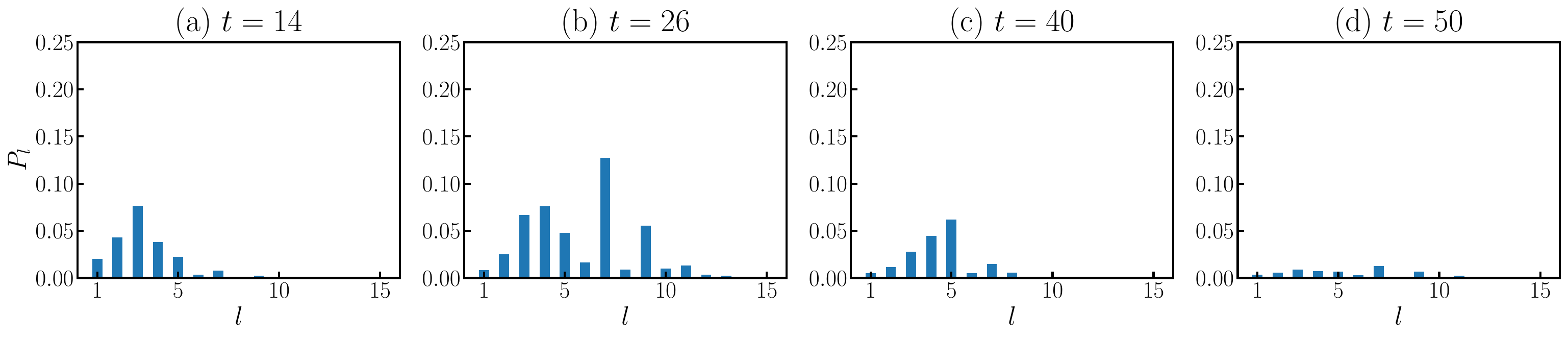}
    \caption{$P_l$ at different times with $m = 0.1, \varepsilon = 0.2, \bar{k} = 6\times(2\pi)/L$, corresponding to the inelastic scattering process shown in the second column of Fig.~\ref{fig: scattering_m0.1}. (a) Time slice before the collision. (b) During the collision, there is an enhanced probability for forming longer strings with $l=10$ compared to Fig.~\ref{fig: strings_m0.1_eps_0.2_k4}, due to the higher momentum in this case. (c) Shortly after the collision the probabilities for having a long flux tube decrease again. (d) When the two mesons are well separated we observe similar situation as for the smaller momentum.}
    \label{fig: strings_m0.1_eps_0.2_k6}
\end{figure}
\begin{figure}[H]
    \centering
    \includegraphics[width=0.9\linewidth]{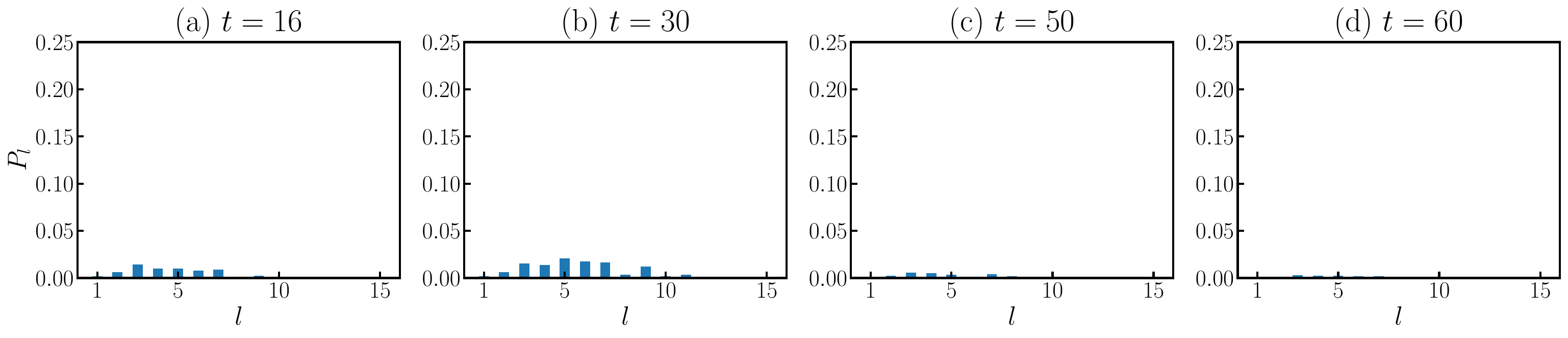}
    \caption{$P_l$ at different times for $m = 0.1$, $\varepsilon = 0.2$, and $\bar{k} = 8 \times (2\pi/L)$, corresponding to the inelastic scattering process shown in the third column of Fig.~\ref{fig: scattering_m0.1}. Compared to the previous two inelastic scattering cases with lower momentum, a smaller probability of forming a single flux string is observed here, indicating a faster hadronization process. (a) Before the collision. (b) During the collision. (c) Shortly after the collision. (d) After the collision when the two mesons are well separated again.}
    \label{fig: strings_m0.1_eps_0.2_k8}
\end{figure}
\begin{figure}[H]
    \centering
    \includegraphics[width=0.9\linewidth]{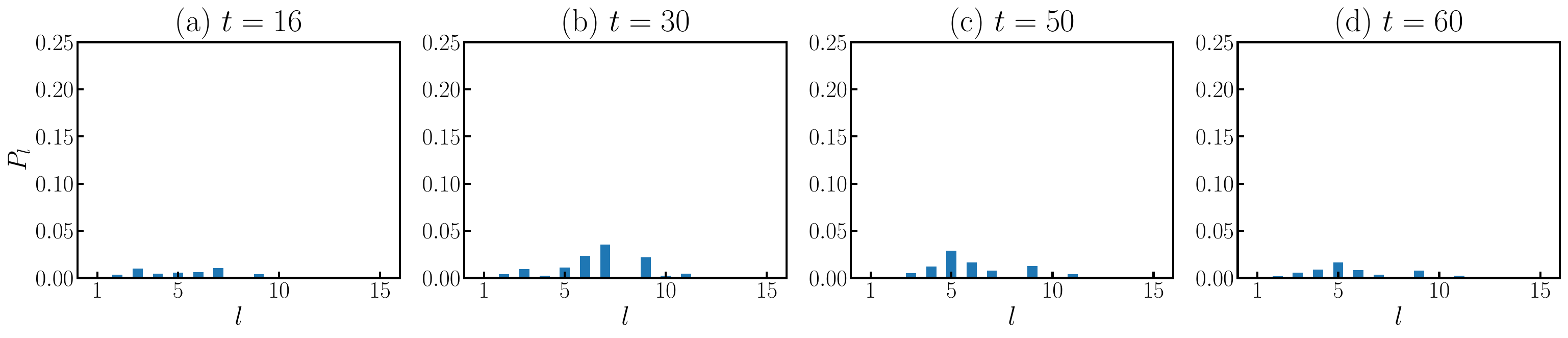}
    \caption{$P_l$ at different times for $m = 0.2$, $\varepsilon = 0.2$, and $\bar{k} = 8 \times (2\pi/L)$, corresponding to the inelastic scattering process shown in the fourth column of Fig.~\ref{fig: scattering_m0.1}. (a) Before the collision. (b) During the collision. (c) Shortly after the collision. (d) After the collision when the two mesons are well separated.}
    \label{fig: strings_m0.2_eps_0.2_k8}
\end{figure}

\section{MPS simulation details}\label{app: MPS_detals}
In this work, we use MPS for our classical numerical simulations. For a system with open boundary conditions and $N$ sites, the ansatz is given by~\cite{Schollwoeck2011,Orus2014a,Banuls2023}
\begin{align}
    \ket{\psi} = \sum_{i_1,i_2,\dots,i_N}^d A_1^{i_1}A_2^{i_2}\dots A_N^{i_N}\ket{i_1}\otimes\ket{i_2}\otimes\dots \otimes\ket{i_N},
    \label{eq:MPS}
\end{align}
where $d$ is the local dimension of the Hilbert space on every site and $\ket{i_k}_{k=1}^d$ is a local basis for the Hilbert space at site $k$. The quantities $A_k^{i_k}$ are complex $D\times D $ matrices for $1<k<N$ and $A_1^{i_1}$ ($A_N^{i_N}$) is a complex $D$-dimensional row (column) vector. The parameter $D$, which determines the number of parameters in the ansatz, is called the bond dimension of the MPS and limits the maximum von Neumann entropy in the state, $S\leq \log_2D$, which is in turn a measure for the amount of quantum correlations. For our numerical simulations, we use the ITensor library~\cite{Fishman_2022,Fishman2022b}. Note that while we are using periodic boundary conditions for our Hamiltonian, we nevertheless use MPS with open boundary conditions. This introduces long-range terms, which connect the first and the last site in the MPS, and can potentially lead to long-range correlations requiring a larger value of $D$ to be captured. This decision is motivated by the fact, that MPS algorithms using open boundary are in general more efficent and stable than their counterparts for periodic boundary conditions~\cite{Schollwoeck2011}.

More specifically, for computing ground states and excitations, we use standard variational methods and optimize the tensors iteratively such that the energy expectation value is minimized. For constructing the basis required for the QSE  and the simulation of time evolution, we use block decimation methods, where we apply local operators to the MPS and reduce the bond dimension of the resulting state by performing a singular value decomposition and discarding all singular values which are smaller than a given threshold, also referred to as cutoff. For the QSE used to construct meson creation operators and to generate eigenstates, we use a cutoff of $10^{-10}$. When simulating the scattering dynamics, we set the cutoff to $10^{-9}$ to control the growth of the bond dimension $D$, which reaches maximum values of approximately $1000$ in the most challenging setting. Additionally, for calculating the meson numbers defined in Eq.~\eqref{eq: meson_number}, we use a larger cutoff of $10^{-8}$, because we need to calculate the expectation of nonlocal operators in the state with large bond dimension after the collision.

Furthermore, when simulating time evolution, we use the second order Trotterization of the time evolution operator, given by
\begin{equation}
    e^{-iHt} \approx \left( e^{-i H_{\mathrm{odd}} \ \frac{\Delta t}{2}} e^{-i H_{\mathrm{even}} \ \Delta t }  e^{-i H_{\mathrm{odd}} \ \frac{\Delta t}{2}} \right)^{t/\Delta t},
\end{equation}
where $\Delta t$ is the time step and $e^{-iH_{\mathrm{even}} \ \Delta t}$ ($e^{-iH_{\mathrm{odd}} \ \Delta t}$) represents the Hamiltonian terms starting at even (odd) matter sites given by
\begin{equation}
    e^{-iH_{\mathrm{even}} \  \Delta t} = \prod_{n\in \mathrm{even}} e^{-iH_n \Delta t}
\end{equation}
for even sites and analogously for the odd sites. Specifically for $n < L$, $H_n$ reads
\begin{equation}    
    H_n = -\frac{i}{2a} \left(\sigma_n^{+} Z_{g, n} \sigma_{n+1}^- - \text{h.c.}\right)+ \frac{m}{4}(-1)^n \left( \sigma_n^z - \sigma_{n+1}^z \right) + \varepsilon X_{g, n},
\end{equation}
and for $n=L$
\begin{equation}    
    H_L = \frac{1}{2a} \left(\prod_{l<L}(-i\sigma_l^z) \sigma_L^{+} Z_{g, L} \sigma_{1}^- + \text{h.c.} \right) + \frac{m}{4}(-1)^L \left( \sigma_L^z - \sigma_{1}^z \right) + \varepsilon X_{g, L},
\end{equation}
which takes into account the periodic boundary conditions.

\twocolumngrid
\bibliography{references}
\end{document}